# Local Graph Clustering Beyond Cheeger's Inequality*

Zeyuan Allen Zhu
zeyuan@csail.mit.edu
MIT CSAIL

Silvio Lattanzi
silviol@google.com
Google Research

Vahab Mirrokni
mirrokni@google.com
Google Research

November 3, 2013

**Abstract**

Motivated by applications of large-scale graph clustering, we study random-walk-based *local* algorithms whose running times depend only on the size of the output cluster, rather than the entire graph. All previously known such algorithms guarantee an output conductance of $\tilde{O}(\sqrt{\phi(A)})$ when the target set $A$ has conductance $\phi(A) \in [0,1]$. In this paper, we improve it to

$$\tilde{O}\bigg( \min\Big\{ \sqrt{\phi(A)}, \frac{\phi(A)}{\sqrt{\mathsf{Conn}(A)}} \Big\} \bigg) ,$$

where the internal connectivity parameter $\mathsf{Conn}(A) \in [0,1]$ is defined as the reciprocal of the mixing time of the random walk over the induced subgraph on $A$.

For instance, using $\mathsf{Conn}(A) = \Omega(\lambda(A)/\log n)$ where $\lambda$ is the second eigenvalue of the Laplacian of the induced subgraph on $A$, our conductance guarantee can be as good as $\tilde{O}(\phi(A)/\sqrt{\lambda(A)})$. This builds an interesting connection to the recent advance of the so-called *improved Cheeger's Inequality* [KLL+13], which says that global spectral algorithms can provide a conductance guarantee of $O(\phi_{\mathsf{opt}}/\sqrt{\lambda_3})$ instead of $O(\sqrt{\phi_{\mathsf{opt}}})$.

In addition, we provide theoretical guarantee on the clustering accuracy (in terms of precision and recall) of the output set. We also prove that our analysis is tight, and perform empirical evaluation to support our theory on both synthetic and real data.

It is worth noting that, our analysis outperforms prior work when the cluster is *well-connected*. In fact, the better it is well-connected inside, the more significant improvement (both in terms of conductance and accuracy) we can obtain. Our results shed light on why in practice some random-walk-based algorithms perform better than its previous theory, and help guide future research about local clustering.

## 1 Introduction

As a central problem in machine learning, clustering methods have been applied to data mining, computer vision, social network analysis. Although a huge number of results are known in this area, there is still need to explore methods that are robust and efficient on large data sets, and have good theoretical guarantees. In particular, several algorithms restrict the number of clusters, or impose constraints that make these algorithms impractical for large data sets.

To solve those issues, recently, local random-walk clustering algorithms [ST04, ACL06, ST13, AP09, OT12] have been introduced. The main idea behind those algorithms is to find a good cluster around a specific node. These techniques, thanks to their scalability, has had high impact in practical applications [LLDM09, GLMY11, GS12, AGM12, LLM10, WLS+12]. Nevertheless, the theoretical understanding of these techniques is still very limited. In this paper, we make an important contribution in this direction. First, we relate for the first time the performance of these local algorithms with the *internal connectivity* of a cluster instead of analyzing only its external connectivity. This change of perspective is relevant for practical applications where we are not only interested to find clusters that are loosely connected with the rest of the world, but also clusters that are well-connected internally. In particular, we show theoretically and empirically that this internal connectivity is a fundamental parameter for those algorithms and, by leveraging it, it is possible to improve their performances.

Formally, we study the clustering problem where the data set is given by a similarity matrix as a graph: given an undirected[1] graph $G = (V, E)$, we want to find a set $S$ that minimizes the relative number of edges going out of $S$ with respect to the size of $S$ (or the size of $\bar{S}$ if $S$ is larger than $\bar{S}$). To capture this concept rigorously, we consider the *(cut) conductance* of a set $S$ as:[2]

$$\phi(S) \stackrel{\text{def}}{=} \frac{|E(S,\bar{S})|}{\min\{\mathrm{vol}(S), \mathrm{vol}(\bar{S})\}} \ ,$$

where $\mathrm{vol}(S) \stackrel{\text{def}}{=} \sum_{v \in S} \deg(v)$. Finding $S$ with the smallest $\phi(S)$ is called the conductance minimization. This measure is a well-studied measure in different disciplines [SM00, ST04, ACL06, GLMY11, GS12], and has been identified as one of the most important cut-based measures in the literature [Sch07]. Many approximation algorithms have been developed for the problem, but most of them are global ones: their running time depends at least linearly on the size of the graph. A recent trend, initiated by Spielman and Teng [ST04], and then followed by [ST13, ACL06, AP09, OT12], attempts to solve this conductance minimization problem *locally*, with running time only dependent on the volume of the output set.

In particular, if there exists a set $A \subset V$ with conductance $\phi(A)$, these local algorithms guarantee the existence of some set $A^g \subseteq A$ with at least half the volume, such that for any "good" starting vertex $v \in A^g$, they output a set $S$ with conductance $\phi(S) = \tilde{O}(\sqrt{\phi(A)})$.

## 1.1 The Internal Connectivity of a Cluster

All local clustering algorithms developed so far, both theoretical ones and empirical ones, only assume that $\phi(A)$ is small, i.e., $A$ is poorly connected to $\bar{A}$. Notice that such set $A$, no matter how small $\phi(A)$ is, may be poorly connected or even disconnected inside. This cannot happen in reality if $A$ is a "good" cluster, and in practice we are often interested in finding mostly good clusters. This motivates us to study an extra measure on $A$, that is the connectivity of $A$, defined as

$$\mathsf{Conn}(A) \stackrel{\text{def}}{=} \frac{1}{\tau_{\mathrm{mix}}(A)} \in [0,1] \ ,$$

where $\tau_{\mathrm{mix}}(A)$ is the mixing time for a random walk on the subgraph induced by $A$. We will formalize the definition of $\tau_{\mathrm{mix}}(A)$ as well as provide alternative definitions to $\mathsf{Conn}(A)$ in Section 2. It is worth noting here that one can for instance replace $\mathsf{Conn}(A)$ with $\mathsf{Conn}(A) \stackrel{\text{def}}{=} \frac{\lambda(A)}{\log \mathrm{vol}(A)}$ where $\lambda(A)$ is the *spectral gap*, i.e., 1 minus the second largest eigenvalue of the random walk matrix on $G[A]$.

[1] All our results can be easily generalized to weighted graphs.

[2] Others also study related notions such as *normalized cut* or expansion, e.g., $\frac{|E(S,\bar{S})|}{\min\{|S|,|\bar{S}|\}}$ or $\frac{|E(S,\bar{S})|}{|S|\cdot|\bar{S}|}$; there exist well-known reductions between the approximation algorithms on them.

## 1.2 Local Clustering for Finding Well-Connected Clusters

In this paper we assume that, in addition to prior work, the cluster $A$ is *well-connected* and satisfies the following gap assumption

$$\mathsf{Gap}(A) \stackrel{\text{def}}{=} \frac{\mathsf{Conn}(A)}{\phi(A)} \geq \Omega(1) \quad ,$$

which says that $A$ is better connected inside than it is connected to $\bar{A}$. This assumption makes sense in real world scenarios for two main reasons. First, in practice we are often interested in retrieving clusters that have a better connectivity within themselves than with the rest of the graph. Second, in several applications the edges of the graph represent pairwise similarity scores extracted from a machine learning algorithm and so we would expect similar nodes to be well connected within themselves while dissimilar nodes to be loosely connected. As a result, it is not surprising that the notion of connectivity is not new. For instance [KVV04] studied a bicriteria optimization for this objective. However, local algorithms based on the above gap assumption is not well studied.[3]

**Our Positive Result.** Under the gap assumption $\mathsf{Gap}(A) \geq \Omega(1)$, can we guarantee any better conductance than the previously shown $\tilde{O}(\sqrt{\phi(A)})$ ones? We prove that the answer is affirmative, along with theoretical guarantees on the accuracy of the output cluster. In particular, we prove:

**Theorem 1.** *There exists a constant $c > 0$ such that, for any non-empty set $A \subset V$ with $\mathsf{Gap}(A) \geq c$, there exists some $A^g \subseteq A$ with $\mathrm{vol}(A^g) \geq \frac{1}{2}\mathrm{vol}(A)$ such that, when choosing a starting vertex $v \in A^g$, the* `PageRank-Nibble` *algorithm outputs a set $S$ with*

1. $\mathrm{vol}(S \setminus A) \leq O\big(\frac{\phi(A)}{\mathsf{Conn}(A)}\big) \cdot \mathrm{vol}(A) = O\big(\frac{1}{\mathsf{Gap}(A)}\big) \cdot \mathrm{vol}(A)$,
2. $\mathrm{vol}(A \setminus S) \leq O\big(\frac{\phi(A)}{\mathsf{Conn}(A)}\big) \cdot \mathrm{vol}(A) = O\big(\frac{1}{\mathsf{Gap}(A)}\big) \cdot \mathrm{vol}(A)$,
3. $\phi(S) \leq O\left(\frac{\phi(A)}{\sqrt{\mathsf{Conn}(A)}}\right) = O\left(\frac{\sqrt{\phi(A)}}{\sqrt{\mathsf{Gap}(A)}}\right)$, *and*

*with running time $O\big(\frac{\mathrm{vol}(A)}{\mathsf{Conn}(A)}\big) \leq O\big(\frac{\mathrm{vol}(A)}{\phi(A)}\big)$.*

We interpret the above theorem as follows. The first two properties imply that under $\mathsf{Gap}(A) \geq \Omega(1)$, the volume for $\mathrm{vol}(S \setminus A)$ and $\mathrm{vol}(A \setminus S)$ are both small in comparison to $\mathrm{vol}(A)$, and the larger the gap is, the more accurate $S$ approximates $A$.[4] For the third property on the conductance $\phi(S)$, we notice that our guarantee $O(\sqrt{\phi(A)/\mathsf{Gap}(A)}) \leq O(\sqrt{\phi(A)})$ outperforms all previous work on local clustering under this parameter regime. In addition, $\mathsf{Gap}(A)$ might be very large in reality. For instance when $A$ is a very-well-connected cluster it might satisfy $\mathsf{Conn}(A) = \mathsf{polylog}(n)$. In this case our Theorem 1 guarantees a $\mathsf{polylog}(n)$ true-approximation to the conductance.

Our proof of Theorem 1 uses almost the same PageRank algorithm as [ACL06], but with a very different analysis specifically designed for our gap assumption.[5] This algorithm is simple and clean, and can be described in four steps: 1) compute the (approximate) PageRank vector starting from a vertex $v \in A^g$ with carefully chosen parameters, 2) sort all the vertices according to their (normalized) probabilities in this vector, 3) study all *sweep cuts* that are those separating high-value vertices from low-value ones, and 4) output the sweep cut with the best conductance. See Algorithm 1 on page 18 for details.

[3] One relevant paper using this assumption is [MMV12], who provided a *global* SDP-based algorithm to approximate the conductance.

[4] Very recently, [WLS+12] studied a variant of the PageRank random walk and their first experiment —although analyzed in a different perspective— essentially confirmed our first two properties in Theorem 1. However, they have not attempted to explain this in theory.

[5] Interestingly, their theorems do not imply any new result in our setting at least in any obvious way, and thus proofs different from the previous work are necessary in this paper. To the best of our knowledge, equation (3.1) is the only part that is a consequence of their result, and we will mention it without proof.

**An Unconditional Result.** In reality one may find it hard to check if the assumption $\mathsf{Gap}(A) \geq \Omega(1)$ is satisfied, and thus we state a simple corollary to the above theorem without this assumption. Note that some

**Corollary 2.** *For any non-empty set $A \subset V$, there exists some $A^g \subseteq A$ with $\mathrm{vol}(A^g) \geq \frac{1}{2}\mathrm{vol}(A)$ such that, when choosing a starting vertex $v \in A^g$, the* `PageRank-Nibble` *algorithm runs in time $O\big(\frac{\mathrm{vol}(A)}{\phi(A)}\big)$ and outputs a set $S$ with*

$$\phi(S) \leq \begin{cases} O\big(\sqrt{\phi(A) \cdot \log \mathrm{vol}(A)}\big), & \text{if } \mathsf{Conn}(A) < c \cdot \phi(A); \\ O\big(\phi(A)/\sqrt{\mathsf{Conn}(A)}\big), & \text{if } \mathsf{Conn}(A) \geq c \cdot \phi(A). \end{cases}$$

*Or more briefly:*

$$\phi(S) \leq \tilde{O}\Bigg( \min\Big\{ \sqrt{\phi(A)}, \frac{\phi(A)}{\sqrt{\mathsf{Conn}(A)}} \Big\} \Bigg) .$$

*Recall that one can choose $\mathsf{Conn}(A) = 1/\tau_{\mathrm{mix}}(A)$ or $\mathsf{Conn}(A) = \lambda(A)/\log \mathrm{vol}(A)$.*

The proof to the above corollary is straightforward. One can simply study two different analyses of the same algorithm `PageRank-Nibble` (with slightly different parameters): one is ours, which only works when $\mathsf{Gap}(A) \geq c$; and the other one is the original analysis of Andersen, Chung and Lang [ACL06], which guarantees an output conductance of $O(\sqrt{\phi(A) \log \mathrm{vol}(A)})$ in the same running time.

**Connection to the Improved Cheeger's Inequality.** Almost simultaneous to the appearance of the first version of this paper [ZLM13], Kwok et al. [KLL+13] discover independently a similar behavior to our result but on global and spectral algorithms, which they call *the improved Cheeger's Inequality.* Let $\phi_{\mathsf{opt}}$ be the optimal conductance of $G$, and $v$ the second eigenvector of the normalized Laplacian matrix of $G$. Using Cheeger's Inequality, one can show that the best sweep cut on $v$ provides a conductance of $O(\sqrt{\phi_{\mathsf{opt}}})$; the improved Cheeger's Inequality says that the conductance guarantee can be improved to $O\left(\frac{\phi_{\mathsf{opt}}}{\sqrt{\lambda_3}}\right)$ where $\lambda_3$ is the third smallest eigenvalue. In other words, the performance (for the same algorithm) is improved when for instance both sides of the desired cut are well-connected (e.g., expanders). Our Theorem 1 and Corollary 2 show that this same behavior occurs for random-walk based local algorithms.

## 1.3 Tightness of Our Analysis

We also prove that our analysis is tight.

**Theorem 3.** *For any constant $c > 0$, there exists a family of graphs $G = (V, E)$ and a non-empty $A \subset V$ with $\mathsf{Gap}(A) > c$, such that for all starting vertices $v \in A$, none of the sweep-cut based algorithm on the PageRank vector can output a set $S$ with conductance better than $O(\phi(A)/\sqrt{\mathsf{Conn}(A)})$.*

We prove this tightness result by illustrating a hard instance, and proving upper and lower bounds on the probabilities of reaching specific vertices (up to a very high precision). In fact, even the description of the hard instance is somewhat non-trivial and different from the improved Cheeger's Inequality case where the hard instance simply a cycle.

Although Theorem 3 does not fully rule out the existence of another local algorithm that can perform better than $O(\phi(A)/\sqrt{\mathsf{Conn}(A)})$, we conjecture that all existing random-walk-based local clustering algorithms share this same hard instance and cannot outperform $O(\phi(A)/\sqrt{\mathsf{Conn}(A)})$. This is analogous to the classical case (without the connectivity assumption) where all existing local algorithms provide $\tilde{O}(\sqrt{\phi(A)})$ due to Cheeger's inequality.

In the first version of this paper [ZLM13], we raised as an interesting open question to design a flow-based local algorithm to overcome this barrier under our connectivity assumption $\mathsf{Gap}(A) \geq \Omega(1)$. Lately, Orecchia and Zhu have made this possible and obtained an $O(1)$-approximation to conductance under this same assumption [OZ14]. Their result is built on ours: it requires a preliminary run of the `PageRank-Nibble` algorithm, the use of our better analysis, and a non-trivial localization of the *cut-improvement algorithm* from the seminar work of Andersen and Lang [AL08]. It is worth pointing out that they achieve this better conductance approximation at the expense of losing the accuracy guarantee of the output cluster (see the first two items of our Theorem 1).

### 1.4 Prior Work

Most relevant to our work are the ones on local algorithms for clustering. After the first such result [ST04, ST13], Andersen, Chung and Lang [ACL06] simply compute a Pagerank random walk vector and then show that one of its sweep cuts satisfies conductance $O(\sqrt{\phi(A)\log\mathrm{vol}(A)})$. The computation of this Pagerank vector is deterministic and is essentially the algorithm we adopt in this paper. [AP09, OT12] use the theory of evolving set from [MP03]. They study a stochastic volume-biased evolving set process that is similar to a random work. This leads to a better (but probabilistic) running time and but essentially with the same conductance guarantee.

The problem of conductance minimization is UGC-hard to approximate within any constant factor [CKK+06]. On the positive side, spectral partitioning algorithms output a solution with conductance $O(\sqrt{\phi_{\mathsf{opt}}})$ where this idea traces back to [Alo86] and [SJ89]; Leighton and Rao [LR99] provide a first $O(\log n)$ approximation; and Arora, Rao and Vazirani [ARV09] provide a $O(\sqrt{\log n})$ approximation. Those results, along with recent improvements on the running time by for instance [OSV12, OSVV08, AHK10, AK07, She09], are all *global* algorithms: their time complexities depend at least linearly on the size of $G$. There are also work in machine learning to make such global algorithms practical, including the work of [LC10] for spectral partitioning.

Less relevant to our work are supervised learning on finding clusters, and there exist algorithms that have a sub-linear running time in terms of the size of the training set [ZCZ+09, SS08].

On the empirical side, random-walk-based graph clustering algorithms have been widely used in practice [GS12, GLMY11, ACE+13, AGM12] as they can be implemented in a distributed manner for very big graphs using map-reduce or similar distributed graph mining algorithms [LLDM09, GLMY11, GS12, AGM12]. Such local algorithms have been applied for (overlapping) clustering of big graphs for distributed computation [AGM12], or community detection on huge Youtube video graphs [GLMY11]. There also exist variants of the random walk, such as the multi-agent random walk, that are known to be local and perform well in practice [AvL10].

More recently, [WLS+12] studied a slight variant of the PageRank random walk and performed supportive experiments on it. Their experiments confirmed the first two properties in our Theorem 1, but their theoretical results are not strong enough to confirm it. This is because there is no well-connectedness assumption in their paper so they are forced to study random walks that start from a random vertex selected in $A$, rather than a fixed one like ours. In addition, they have not argued about the conductance (like our third property in Theorem 1) of the set they output.

Clustering is an important technique for community detections, and indeed local clustering algorithms have been widely applied there, see for instance [AL06]. Sometimes researchers care about finding all communities, i.e., clusters, in the entire graph and this can be done by repeatedly applying local clustering algorithms. However, if the ultimate goal is to find all clusters, global algorithms perform better in at least in terms of minimizing conductance [LLDM09, GLMY11, GS12, AGM12, LLM10].

### 1.5 Roadmap

We provide necessary preliminaries in Section 2, and they are followed by the high level ideas for the proofs (as long as the actual proofs) for Theorem 1 in Section 3 and Section 4. We then briefly describe how to prove our tightness result in Section 5 while deferring the analysis to Appendix A, and end this paper with empirical studies in Section 6. In Appendix B we briefly summarize and show some property for the algorithm `Approximate-PR` of Andersen, Chung and Lang for completeness.

## 2 Preliminaries

### 2.1 Problem Formulation

Consider an undirected graph $G(V,E)$ with $n = |V|$ vertices and $m = |E|$ edges. For any vertex $u \in V$ the degree of $u$ is denoted by $\deg(u)$, and for any subset of the vertices $S \subseteq V$, *volume* of $S$ is denoted by $\mathrm{vol}(S) \stackrel{\mathrm{def}}{=} \sum_{u\in S}\deg(u)$. Given two subsets $A, B \subset V$, let $E(A,B)$ be the set of edges between $A$ and $B$.

For a vertex set $S \subseteq V$, we denote by $G[S]$ the induced subgraph of $G$ on $S$ with outgoing edges removed, by $\deg_S(u)$ the degree of vertex $u \in S$ in $G[S]$, and by $\mathrm{vol}_S(T)$ the volume of $T \subseteq S$ in $G[S]$.

We respectively define the *(cut) conductance* and the *set conductance* of a non-empty set $S \subseteq V$ as follows:

$$\phi(S) \stackrel{\mathrm{def}}{=} \frac{|E(S,\bar{S})|}{\min\{\mathrm{vol}(S),\mathrm{vol}(\bar{S})\}} ,$$

$$\phi_{\mathsf{s}}(S) \stackrel{\mathrm{def}}{=} \min_{\emptyset\subset T\subset S} \frac{|E(T, S\setminus T)|}{\min\{\mathrm{vol}_S(T),\mathrm{vol}_S(S\setminus T)\}} .$$

Here $\phi_{\mathsf{s}}(S)$ is classically known as the conductance of $S$ on the induced subgraph $G[S]$.

We formalize our goal in this paper as a *promise problem*. Specifically, we assume the existence of a non-empty target cluster of the vertices $A \subset V$ satisfying $\mathrm{vol}(A) \le \frac{1}{2}\mathrm{vol}(V)$. This set $A$ is *not* known to the algorithm. The goal is to find some set $S$ that "reasonably" approximates $A$, and at the same time be *local*: running in time proportional to $\mathrm{vol}(A)$ rather than $n$ or $m$.

**Our assumption.** We assume that the target set $A$ is *well-connected*, i.e., the following gap assumption:

$$\mathsf{Gap}(A) \stackrel{\mathrm{def}}{=} \frac{\mathsf{Conn}(A)}{\phi(A)} \stackrel{\mathrm{def}}{=} \frac{1/\tau_{\mathrm{mix}}(A)}{\phi(A)} \ge \Omega(1) \qquad \text{(Gap Assumption)}$$

holds throughout this paper. This assumption can be understood as the cluster $A$ is more well-connected inside than it is connected to $\bar{A}$. For all the positive results of this paper, one can replace this assumption with

$$\mathsf{Gap}(A) = \frac{\mathsf{Conn}(A)}{\phi(A)} \stackrel{\mathrm{def}}{=} \frac{\lambda(A)/\log\mathrm{vol}(A)}{\phi(A)} \ge \Omega(1) , \quad \text{or} \qquad \text{(Gap Assumption')}$$

$$\mathsf{Gap}(A) = \frac{\mathsf{Conn}(A)}{\phi(A)} \stackrel{\mathrm{def}}{=} \frac{\phi_{\mathsf{s}}^2(A)/\log\mathrm{vol}(A)}{\phi(A)} \ge \Omega(1) \qquad \text{(Gap Assumption'')}$$

- Here $\lambda(A)$ is the *spectral gap*, that is the difference between the first and second largest eigenvalues of the lazy random walk matrix on $G[A]$. (Notice that the largest eigenvalue of any random walk matrix is always 1.) Equivalently, $\lambda(A)$ can be defined as the second smallest eigenvalue of the Laplacian matrix of $G[A]$.
- Here $\tau_{\mathrm{mix}}$ is the *mixing time for the relative pointwise distance* in $G[A]$ (cf. Definition 6.14 in [MR95]), that is, the minimum time required for a lazy random walk to mix *relatively* on all vertices regardless of the starting distribution. Formally, let $W_A$ be the lazy random walk matrix on $G[A]$, and $\pi$ be the stationary distribution on $G[A]$ that is $\pi(u) = \deg_A(u)/\mathrm{vol}_A(A)$, then
$$\tau_{\mathrm{mix}} = \min\left\{ t \in \mathbb{Z}_{\geq 0} \ : \ \max_{u,v} \left| \frac{(\chi_v W_A^t)(u) - \pi(u)}{\pi(u)} \right| \leq \frac{1}{2} \right\} \ .$$

Notice that using Cheeger's inequality, we always have $\frac{\phi_{\mathsf{s}}(A)^2}{\log \mathrm{vol}(A)} \leq O\left(\frac{\lambda(A)}{\log \mathrm{vol}(A)}\right) \leq O\big(\frac{1}{\tau_{\mathrm{mix}}}\big)$. This is why (Gap Assumption) is weaker than (Gap Assumption') which is then weaker than (Gap Assumption").

**Input parameters.** Similar to prior work on local clustering, we assume the algorithm takes as input:

- *Some "good" starting vertex $v \in A$, and an oracle to output the set of neighbors for any given vertex.*

  This requirement is essential because without such an oracle the algorithm may have to read all inputs and cannot be sublinear in time; and without a starting vertex the sublinear-time algorithm may be unable to even find an element in $A$.

  We also need $v$ to be "good", as for instance the vertices on the boundary of $A$ may not be helpful enough in finding good clusters. We call the set of good vertices $A^g \subseteq A$, and a local algorithm needs to ensure that $A^g$ is large, e.g., $\mathrm{vol}(A^g) \geq \frac{1}{2}\mathrm{vol}(A)$. This assumption is unavoidable in all local clustering work. One can replace this $\frac{1}{2}$ by any other constant at the expense of worsening the guarantees by a constant factor.

- *The value of* $\mathsf{Conn}(A)$.

  In practice $\mathsf{Conn}(A)$ can be viewed as a parameter and can be tuned for specific data. This is in contrast to the value of $\phi(A)$ that is the target conductance and does not need to be known by the algorithm. In prior work when $\phi(A)$ is the only quantity studied, $\phi(A)$ plays both roles as a (known) tuning parameter and as a target.

- *A value* $\mathrm{vol}_0$ *satisfying* $\mathrm{vol}(A) \in [\mathrm{vol}_0, 2\mathrm{vol}_0]$.

  This requirement is optional since otherwise the algorithm can try out different powers of 2 and pick the smallest one with a valid output. It blows up the running time only by a constant factor for local algorithms, since the running time of the last trial dominates.

## 2.2 PageRank Random Walk

We use the convention of writing vectors as row vectors in this paper. Let $A$ be the adjacency matrix of $G$, and let $D$ be the diagonal matrix with $D_{ii} = \deg(i)$, then the *lazy random walk matrix* $W \stackrel{\mathrm{def}}{=} \frac{1}{2}(I + D^{-1}A)$. Accordingly, the PageRank vector $pr_{s,\alpha}$, is defined to be the unique solution of the following linear equation (cf. [ACL06]):
$$pr_{s,\alpha} = \alpha s + (1-\alpha) pr_{s,\alpha} W \ ,$$

where $\alpha \in (0,1]$ is the *teleport probability* and $s$ is a *starting vector*. Here $s$ is usually a probability vector: its entries are in $[0,1]$ and sum up to 1. For technical reasons we may use an arbitrary (and possibly negative) vector $s$ inside the proof. When it is clear from the context, we drop $\alpha$ in the subscript for cleanness.

Given a vertex $u \in V$, let $\chi_u \in \{0,1\}^{|V|}$ be the indicator vector that is 1 only at vertex $u$. Given non-empty subset $S \subseteq V$ we denote by $\pi_S$ the degree-normalized uniform distribution on $S$, that is, $\pi_S(u) = \frac{\deg(u)}{\mathrm{vol}(S)}$ when $u \in S$ and 0 otherwise. Very often we study a PageRank vector when $s = \chi_v$ is an indicator vector, and if so we abbreviate $pr_{\chi_v}$ by $pr_v$.

One equivalent way to study $pr_s$ is to imagine the following random procedure: first pick a non-negative integer $t \in \mathbb{Z}_{\geq 0}$ with probability $\alpha(1-\alpha)^t$, then perform a lazy random walk starting at vector $s$ with exactly $t$ steps, and at last define $pr_s$ to be the vector describing the probability of reaching each vertex in this random procedure. In its mathematical formula we have (cf. [Hav02, ACL06]):

**Proposition 2.1.** $pr_s = \alpha s + \alpha \sum_{t=1}^{\infty}(1-\alpha)^t (sW^t)$.

This implies that $pr_s$ is linear: $a \cdot pr_s + b \cdot pr_t = pr_{as+bt}$.

## 2.3 Approximate PageRank Vector

In the seminal work of [ACL06], they defined approximate PageRank vectors and designed an algorithm to compute them efficiently.

**Definition 2.2.** *An* $\varepsilon$-approximate PageRank vector $p$ for $pr_s$ *is a nonnegative PageRank vector* $p = pr_{s-r}$ *where the vector* $r$ *is nonnegative and satisfies* $r(u) \leq \varepsilon \deg(u)$ *for all* $u \in V$.

**Proposition 2.3.** *For any starting vector* $s$ *with* $\|s\|_1 \leq 1$ *and* $\varepsilon \in (0,1]$, *one can compute an* $\varepsilon$*-approximate PageRank vector* $p = pr_{s-r}$ *for some* $r$ *in time* $O\left(\frac{1}{\varepsilon\alpha}\right)$, *with* $\mathrm{vol}(\mathrm{supp}(p)) \leq \frac{2}{(1-\alpha)\varepsilon}$.

For completeness we provide the algorithm and its proof in Appendix B. It can be verified that:

$$\forall u \in V, \quad pr_s(u) \geq p(u) \geq pr_s(u) - \varepsilon \deg(u) \ . \tag{2.1}$$

## 2.4 Sweep Cuts

Given any approximate PageRank vector $p$, the *sweep cut (or threshold cut)* technique is the one to sort all vertices according to their degree-normalized probabilities $\frac{p(u)}{\deg(u)}$, and then study only those cuts that separate high-value vertices from low-value vertices. More specifically, let $v_1, v_2, \ldots, v_n$ be the decreasing order over all vertices with respect to $\frac{p(u)}{\deg(u)}$. Then, define *sweep sets* $S_j^p \stackrel{\mathrm{def}}{=} \{v_1, \ldots, v_j\}$ for each $j \in [n]$, and sweep cuts are the corresponding cuts $(S_j^p, \overline{S_j^p})$. Usually given a vector $p$, one looks for the best cut:

$$\min_{j \in [n-1]} \phi(S_j^p) \ .$$

In almost all the cases, one only needs to enumerate $j$ over $p(v_j) > 0$, so the above sweep cut procedure runs in time $O\big(\mathrm{vol}(\mathrm{supp}(p)) + |\mathrm{supp}(p)| \cdot \log |\mathrm{supp}(p)|\big)$. This running time is dominated by the time to compute $p$ (see Proposition 2.3), so it is negligible.

### 2.5 Lovász-Simonovits Curve

Our proof requires the technique of *Lovász-Simonovits Curve* that has been more or less used in all local clustering algorithms so far. This technique was originally introduced by Lovász and Simonovits [LS90, LS93] to study the mixing rate of Markov chains. In our language, from a probability vector $p$ on vertices, one can introduce a function $p[x]$ on real number $x \in [0, 2m]$. This function $p[x]$ is piecewise linear, and is characterized by all of its end points as follows (letting $p(S) \stackrel{\text{def}}{=} \sum_{a \in S} p(a)$):

$$p[0] \stackrel{\text{def}}{=} 0, \quad p[\text{vol}(S_j^p)] \stackrel{\text{def}}{=} p(S_j^p) \text{ for each } j \in [n] \enspace .$$

In other words, for any $x \in [\text{vol}(S_j^p), \text{vol}(S_{j+1}^p)]$,

$$p[x] \stackrel{\text{def}}{=} p(S_j^p) + \frac{x - \text{vol}(S_j^p)}{\deg(v_{j+1})} p(v_{j+1}) \enspace .$$

Note that $p[x]$ is increasing and concave.

## 3 Our Accuracy Guarantee

In this section, we study PageRank random walks that start at a vertex $v \in A$ with teleport probability $\alpha$. We claim the range of interesting $\alpha$ is $\big[\Omega(\phi(A)), O(\mathsf{Conn}(A))\big]$. This is because, at a high level, when $\alpha \ll \phi(A)$ the random walk will leak too much to $\bar{A}$; while when $\alpha \gg \mathsf{Conn}(A)$ the random walk will not mix well inside $A$. In prior work, $\alpha$ is chosen to be $\Theta(\phi(A))$, and we will adopt the choice of $\alpha = \Theta(\mathsf{Conn}(A)) = \Theta(\phi(A) \cdot \mathsf{Gap}(A))$. Intuitively, this choice of $\alpha$ ensures that under the condition the random walk mixes inside, it makes the walk leak as little as possible to $\bar{A}$. We prove the above intuition rigorously in this section. Specifically, we first show some properties on the exact PageRank vector in Section 3.1, and then move to the approximate vector in Section 3.2. This essentially proves the first two properties of Theorem 1.

### 3.1 Properties on the Exact Vector

We first introduce a new notation $\widetilde{pr}_s$, that is the PageRank vector (with teleport probability $\alpha$) starting at vector $s$ but walking on the subgraph $G[A]$.

Next, we choose the set of "good" starting vertices $A^g$ to satisfy two properties: (1) the total probability of leakage is upper bounded by $\frac{2\phi(A)}{\alpha}$, and (2) $pr_v$ is close to $\widetilde{pr}_v$ for vertices in $A$. Note that the latter implies that $pr_v$ mixes well inside $A$ as long as $\widetilde{pr}_v$ does so.

**Lemma 3.1.** *There exists a set $A^g \subseteq A$ with volume $\text{vol}(A^g) \geq \frac{1}{2}\text{vol}(A)$ such that, for any vertex $v \in A^g$, in a PageRank vector with teleport probability $\alpha$ starting at $v$, we have:*

$$\sum_{u \notin A} pr_v(u) \leq \frac{2\phi(A)}{\alpha} \enspace . \tag{3.1}$$

*In addition, there exists a non-negative* leakage vector $l \in [0,1]^{|V|}$ *with norm* $\|l\|_1 \leq \frac{2\phi(A)}{\alpha}$ *satisfying*

$$\forall u \in A, \quad pr_v(u) \geq \widetilde{pr}_v(u) - \widetilde{pr}_l(u) \enspace . \tag{3.2}$$

(Details of the proof are in Section 3.3.)

*Proof sketch.* The proof for the first property (3.1) is classical and can be found in [ACL06]. The idea is to study an auxiliary PageRank random walk with teleport probability $\alpha$ starting at the degree-normalized uniform distribution $\pi_A$, and by simple computation, this random walk leaks to $\bar{A}$ with probability no more than $\phi(A)/\alpha$. Then, using Markov bound, there exists $A^g \subseteq A$ with $\mathrm{vol}(A^g) \geq \frac{1}{2}\mathrm{vol}(A)$ such that for each starting vertex $v \in A^g$, this leakage is no more than $\frac{2\phi(A)}{\alpha}$. This implies (3.1) immediately.

The interesting part is (3.2). Note that $pr_v$ can be viewed as the probability vector from the following random procedure: start from vertex $v$, then at each step with probability $\alpha$ let the walk stop, and with probability $(1-\alpha)$ follow the matrix $W$ to go to one of its neighbors (or itself) and continue. Now, we divide this procedure into two rounds. In the first round, we run the same PageRank random walk but whenever the walk wants to use an outgoing edge from $A$ to leak, we let it stop and temporarily "hold" this probability mass. We define $l$ to be the non-negative vector where $l(u)$ denotes the amount of probability that we have "held" at vertex $u$. In the second round, we continue our random walk only from vector $l$. It is worth noting that $l$ is non-zero only at boundary vertices in $A$.

Similarly, we divide the PageRank random walk for $\widetilde{pr}_v$ into two rounds. In the first round we hold exactly the same amount of probability $l(u)$ at boundary vertices $u$, and in the second round we start from $l$ but continue this random walk only within $G[A]$. To bound the difference between $pr_v$ and $\widetilde{pr}_v$, we note that they share the same procedure in the first round; while for the second round, the random procedure for $pr_v$ starts at $l$ and walks towards $V \setminus A$ (so in the worst case it may never come back to $A$ again), while that for $\widetilde{pr}_v$ starts at $l$ and walks only inside $G[A]$ so induces a probability vector $\widetilde{pr}_l$ on $A$. This gives (3.2).

At last, to see $\|l\|_1 \leq \frac{2\phi(A)}{\alpha}$, one just needs to verify that $l(u)$ is essentially the probability that the original PageRank random walk leaks from vertex $u$. Then, $\|l\|_1 \leq \frac{2\phi(A)}{\alpha}$ follows from the fact that the total amount of leakage is upper bounded by $\frac{2\phi(A)}{\alpha}$. □

As mentioned earlier, we want to use (3.2) to lower bound $pr_v(u)$ for vertices $u \in A$. We achieve this by first lower bounding $\widetilde{pr}_v$ which is the PageRank random walk on $G[A]$. Given a teleport probability $\alpha$ that is small compared to $\mathsf{Conn}(A)$, this random walk should mix well. We formally state it as the following lemma, and provide its proof in the Section 3.4.

**Lemma 3.2.** *When $\alpha \leq O(\mathsf{Conn}(A))$ we have that*

$$\forall u \in A, \quad \widetilde{pr}_v(u) \geq \frac{4}{5}\frac{\deg_A(u)}{\mathrm{vol}(A)} \ .$$

*Here $\deg_A(u)$ is the degree of $u$ on $G[A]$, but $\mathrm{vol}(A)$ is with respect to the original graph.*

## 3.2 Properties of the Approximate Vector

From this section on we always use $\alpha \leq O(\mathsf{Conn}(A))$. We then fix a starting vertex $v \in A^g$ and study an $\varepsilon$-approximate Pagerank vector for $pr_v$. We choose

$$\varepsilon = \frac{1}{10 \cdot \mathrm{vol}_0} \in \left[\frac{1}{20\mathrm{vol}(A)}, \frac{1}{10\mathrm{vol}(A)}\right] \ . \tag{3.3}$$

For notational simplicity, we denote by $p$ this $\varepsilon$-approximation and recall from Section 2.3 that $p = pr_{\chi_v - r}$ where $r$ is a non-negative vector with $0 \leq r(u) \leq \varepsilon\deg(u)$ for every $u \in V$. Recall from (2.1) that $pr_v(u) \geq p(u) \geq pr_v(u) - \epsilon \cdot \deg(u)$ for all $u \in V$.

We now rewrite Lemma 3.1 in the language of approximate PageRank vectors using Lemma 3.2:

**Corollary 3.3.** *For any $v \in A^g$ and $\alpha \le O(\mathsf{Conn}(A))$, in an $\varepsilon$-approximate PageRank vector to $pr_v$ denoted by $p = pr_{\chi_v - r}$, we have:*

$$\sum_{u \notin A} p(u) \le \frac{2\phi(A)}{\alpha} \text{ and } \sum_{u \notin A} r(u) \le \frac{2\phi(A)}{\alpha} \ .$$

*In addition, there exists a non-negative* leakage vector $l \in [0,1]^V$ *with norm* $\|l\|_1 \le \frac{2\phi(A)}{\alpha}$ *satisfying*

$$\forall u \in A, \quad p(u) \ge \frac{4}{5}\frac{\deg_A(u)}{\mathrm{vol}(A)} - \frac{\deg(u)}{10\mathrm{vol}(A)} - \widetilde{pr}_l(u) \ .$$

*Proof.* The only inequality that requires a proof is $\sum_{u \notin A} r(u) \le \frac{2\phi(A)}{\alpha}$. In fact, if one takes a closer look at the algorithm to compute an approximate Pagerank vector (cf. Appendix B), the total probability mass that will be sent to $r$ on vertices outside $A$, is upper bounded by the probability of leakage. However, the latter is upper bounded by $\frac{2\phi(A)}{\alpha}$ when we choose $A^g$. □

We are now ready to state the main lemma of this section. We show that for all reasonable sweep sets $S$ on this probability vector $p$, it satisfies that $\mathrm{vol}(S \setminus A)$ and $\mathrm{vol}(A \setminus S)$ are both at most $O\big(\frac{\phi(A)}{\alpha}\mathrm{vol}(A)\big)$.

**Lemma 3.4.** *In the same definition of $\alpha$ and $p$ from Corollary 3.3, let sweep set $S_c \stackrel{\text{def}}{=} \big\{u \in V : p(u) \ge c\frac{\deg(u)}{\mathrm{vol}(A)}\big\}$ for any constant $c < \frac{3}{5}$, then we have the following guarantees on the size of $S_c \setminus A$ and $A \setminus S_c$:*

1. $\mathrm{vol}(S_c \setminus A) \le \frac{2\phi(A)}{\alpha c}\mathrm{vol}(A)$, *and*
2. $\mathrm{vol}(A \setminus S_c) \le \Big(\frac{2\phi(A)}{\alpha(\frac{3}{5}-c)} + 8\phi(A)\Big)\mathrm{vol}(A)$.

*Proof.* First we notice that $p(S_c \setminus A) \le p(V \setminus A) \le \frac{2\phi(A)}{\alpha}$ owing to Corollary 3.3, and for each vertex $u \in S_c \setminus A$ it must satisfy $p(u) \ge c\frac{\deg(u)}{\mathrm{vol}(A)}$. Those combined imply $\mathrm{vol}(S_c \setminus A) \le \frac{2\phi(A)}{\alpha c}\mathrm{vol}(A)$ proving the first property.

We show the second property in two steps. First, let $A_b$ be the set of vertices in $A$ such that $\frac{4}{5}\frac{\deg_A(u)}{\mathrm{vol}(A)} - \frac{\deg(u)}{10\mathrm{vol}(A)} < \frac{3}{5}\frac{\deg(u)}{\mathrm{vol}(A)}$. Any such vertex $u \in A_b$ must have $\deg_A(u) < \frac{7}{8}\deg(u)$. This implies that $u$ has to be on the boundary of $A$ and $\mathrm{vol}(A_b) \le 8\phi(A)\mathrm{vol}(A)$.

Next, for a vertex $u \in A \setminus A_b$ we have (using Corollary 3.3 again) $p(u) \ge \frac{3}{5}\frac{\deg(u)}{\mathrm{vol}(A)} - \widetilde{pr}_l(u)$. If we further assume $u \notin S_c$ we have $p(u) < c\frac{\deg(u)}{\mathrm{vol}(A)}$, that implies $\widetilde{pr}_l(u) \ge (\frac{3}{5}-c)\frac{\deg(u)}{\mathrm{vol}(A)}$. As a consequence, the total volume for such vertices (i.e., $\mathrm{vol}(A \setminus (A_b \cup S_c))$) cannot exceed $\frac{\|\widetilde{pr}_l\|_1}{3/5-c}\mathrm{vol}(A)$. At last, we notice that $\widetilde{pr}_l$ is a non-negative probability vector coming from a random walk procedure, so $\|\widetilde{pr}_l\|_1 = \|l\|_1 \le \frac{2\phi(A)}{\alpha}$. This in sum provides that

$$\begin{aligned}\mathrm{vol}(A \setminus S_c) &\le \mathrm{vol}(A \setminus (A_b \cup S_c)) + \mathrm{vol}(A_b) \\ &\le \left(\frac{2\phi(A)}{\alpha(\frac{3}{5}-c)} + 8\phi(A)\right)\mathrm{vol}(A) \ .\end{aligned}$$

□

Note that if one chooses $\alpha = \Theta(\mathsf{Conn}(A))$ in the above lemma, both those two volumes are at most $O(\mathrm{vol}(A)/\mathsf{Gap}(A))$ satisfying the first two properties of Theorem 1.

### 3.3 Proof of Lemma 3.1

**Lemma 3.1.** *There exists a set $A^g \subseteq A$ with volume $\mathrm{vol}(A^g) \geq \frac{1}{2}\mathrm{vol}(A)$ such that, for any vertex $v \in A^g$, in a PageRank vector with teleport probability $\alpha$ starting at $v$, we have:*

$$\sum_{u \notin A} pr_v(u) \leq \frac{2\phi(A)}{\alpha} \quad . \tag{3.1}$$

*In addition, there exists a non-negative* leakage vector $l \in [0,1]^V$ *with norm* $\|l\|_1 \leq \frac{2\phi(A)}{\alpha}$ *satisfying*

$$\forall u \in A, \quad pr_v(u) \geq \widetilde{pr}_v(u) - \widetilde{pr}_l(u) \quad . \tag{3.2}$$

**Leakage event.** We begin our proof by defining the *leaking event* in a random walk procedure. We start the definition of a lazy random walk and then move to a PageRank random walk. At high level, we say that a lazy random walk of length $t$ starting at a vertex $u \in A$ does not *leak* from $A$ if it never goes out of $A$, and let $\mathtt{Leak}(u,t)$ denote the probability that such a random walk leaks.

More formally, for each vertex $u \in V$ in the graph with degree $\deg(u)$, recall that in its random walk graph it actually has degree $2\deg(u)$, with $\deg(u)$ edges going to each of its neighbors, and $\deg(u)$ self-loops. For a vertex $u \in A$, let us call its neighboring edge $(u,v) \in E$ a *bad edge* if $v \notin A$. In addition, if $u$ has $k$ bad edges, we also distinguish $k$ self-loops at $u$ in the lazy random walk graph, and call them *bad self-loops.* Now, we say that a random walk does not leak from $A$, *if it never uses any of those bad edges of self-loops.* The purpose of this definition is to make sure that if a random walk chooses only good edges at each step, it is equivalent to a lazy random walk on the induced subgraph $G[A]$ with outgoing edges removed.

For a PageRank random walk with teleport probability $\alpha$ starting at a vertex $u$, recall that it is also a random procedure and can be viewed as first picking a length $t \in \{0,1,\dots\}$ with probability $\alpha(1-\alpha)^t$, and then performing a lazy random walk of length $t$ starting from $u$. By the linearity of random walk vectors, the probability of leakage for this Pagerank random walk is exactly $\sum_{t=0}^{\infty} \alpha(1-\alpha)^t \mathtt{Leak}(u,t)$.

**Upper bounding leakage.** We now give an upper bound on the probability of leakage. We start with an auxiliary lazy random walk of length $t$ starting from a "uniform" distribution $\pi_A(u)$. Recall that $\pi_A(u) = \frac{\deg(u)}{\mathrm{vol}(A)}$ for $u \in A$ and 0 elsewhere. We now want to show that this random walk leaks with probability at most $1 - t\phi(A)$.[6] This is because, one can verify that: (1) in the first step of this random walk, the probability of leakage is upper bounded by $\phi(A)$ by the definition of conductance; and (2) in the $i$-th step in general, this random walk satisfies $(\pi_A W^{i-1})(u) \leq \pi_A(u)$ for any vertex $u \in A$, and therefore the probability of leakage in the $i$-th step is upper bounded by that in the first step. In sum, the total leakage is at most $t\phi(A)$, or equivalently, $\sum_{u \in A} \pi_A(u)\mathtt{Leak}(u,t) \leq t\phi(A)$.

We now sum this up over the distribution of $t$ in a PageRank random walk:

$$\begin{aligned}\sum_{u \in A} \pi_A(u) \left( \sum_{t=0}^{\infty} \alpha(1-\alpha)^t \mathtt{Leak}(u,t) \right) &= \sum_{t=0}^{\infty} \alpha(1-\alpha)^t \left( \sum_{u \in A} \pi_A(u) \mathtt{Leak}(u,t) \right) \\ &\leq \sum_{t=0}^{\infty} \alpha(1-\alpha)^t t\phi(A) = \frac{\phi(A)(1-\alpha)}{\alpha} \quad .\end{aligned}$$

[6] Note that this step of the proof coincides with that of Proposition 2.5 from [ST13]. Our $t\phi(A)$ is off by a factor of 2 from theirs because we also regard bad self-loops as edges that leak.

This implies, using Markov bound, there exists a set $A^g \subseteq A$ with volume $\mathrm{vol}(A^g) \geq \frac{1}{2}\mathrm{vol}(A)$ satisfying

$$\forall v \in A^g, \qquad \sum_{t=0}^{\infty} \alpha(1-\alpha)^t \mathtt{Leak}(v,t) \leq \frac{2\phi(A)(1-\alpha)}{\alpha} < \frac{2\phi(A)}{\alpha} \ , \tag{3.4}$$

or in words: the probability of leakage is at most $\frac{2\phi(A)(1-\alpha)}{\alpha}$ in a Pagerank random walk that starts at vertex $v \in A^g$. This inequality immediately implies (3.1), so for the rest of the proof, we concentrate on (3.2).

**Lower bounding *pr*.** Now we pick some $v \in A^g$, and try to lower bound $pr_v$. To begin with, we define two $|A| \times |A|$ lazy random walk matrices on the induced subgraph $G[A]$ (recall that $\deg(u)$ is the degree of a vertex and for $u \in A$ we denote by $\deg_A(u)$ the number of neighbors of $u$ inside $A$):

1. Matrix $\widehat{W}$. This is a random walk matrix assuming that all outgoing edges from $A$ being "phantom", that is, at each vertex $u \in A$:
   - it picks each neighbor in $A$ with probability $\frac{1}{2\deg(u)}$, and
   - it stays where it is with probability $\frac{\deg_A(u)}{2\deg(u)}$.

   For instance, let $u$ be a vertex in $A$ with four neighbors $w_1, w_2, w_3, w_4$ such that $w_1, w_2, w_3 \in A$ but $w_4 \notin A$. Then, for a lazy random walk using matrix $\widehat{W}$, if it starts from $u$ then in the next step it stays at $u$ with probability $3/8$, and goes to $w_1, w_2$ and $w_3$ each with probability $1/8$. Note that, for the rest $1/4$ probability (which corresponds to $w_4$) it goes *nowhere* and this random walk "disappears"! This can be viewed as that the random walk leaks $A$.

2. Matrix $\widetilde{W}$. This is a random walk matrix assuming that all outgoing edges from $A$ are removed, that is, at each vertex $u \in A$:
   - it picks each neighbor in $A$ with probability $\frac{1}{2\deg_A(u)}$, and
   - it stays where it is with probability $\frac{1}{2}$.

The major difference between $\widetilde{W}$ and $\widehat{W}$ is that they are normalized by different degrees in the rows, and the rows of $\widetilde{W}$ sum up to 1 but those of $\widehat{W}$ do not necessarily. More specifically, if we denote by $D$ the diagonal matrix with $\deg(u)$ on the diagonal for each vertex $u \in A$, and $D_A$ the diagonal matrix with $\deg_A(u)$ on the diagonal, then $\widehat{W} = D^{-1}D_A\widetilde{W}$. It is worth noting that, if one sums up all entries of the nonnegative vector $\chi_v\widehat{W}^t$, the summation is exactly $1 - \mathtt{Leak}(v,t)$ by our definition of $\mathtt{Leak}$.

We now precisely study the difference between $\widetilde{W}$ and $\widehat{W}$ using the following claim.

**Claim 3.5.** *There exists non-negative vectors $l_t$ for all $t \in \{1,2,\dots\}$ satisfying:*

$$\|l_t\|_1 = \mathtt{Leak}(v,t) - \mathtt{Leak}(v,t-1) \ ,$$

*and*

$$\chi_v\widehat{W}^t = \left(\chi_v\widehat{W}^{t-1} - l_t\right)\widetilde{W} \ .$$

*Proof.* To obtain the result of this claim, we write

$$\begin{aligned}\chi_v \widehat{W}^t &= \left(\chi_v \widehat{W}^{t-1}\right) D^{-1} D_A \widetilde{W} \\ &= \left(\chi_v \widehat{W}^{t-1}\right) \widetilde{W} - \left(\chi_v \widehat{W}^{t-1}\right) (I - D^{-1} D_A) \widetilde{W}\end{aligned}$$

Now, we simply let $l_t \stackrel{\text{def}}{=} \left(\chi_v \widehat{W}^{t-1}\right)(I - D^{-1}D_A)$. It is a non-negative vector because $\deg_A(u)$ is no larger than $\deg(u)$ for all $u \in A$. Furthermore, recall that in the lazy random walk characterized by $\widehat{W}$, the amount of probability to disappear at a vertex $u$ in the $t$-th step, is exactly its probability after a $(t-1)$-th step random walk, i.e., $(\chi_v \widehat{W}^{t-1})(u)$, multiplied by the probability to leak in this step, i.e., $1 - \frac{\deg_A(u)}{\deg(u)}$. Therefore, $l_t(u)$ exactly equals to the amount of probability to disappear in the $t$-th step; or equivalently, $\|l_t\|_1 = \mathtt{Leak}(v,t) - \mathtt{Leak}(v,t-1)$. □

Now we use the above definition of $l_t$ and deduce that:

**Claim 3.6.** *Letting $l \stackrel{\text{def}}{=} \sum_{j=1}^{\infty}(1-\alpha)^{j-1} l_j$, we have $\|l\|_1 \le \frac{2\phi(A)}{\alpha}$ and the following inequality on vector holds coordinate-wisely on all vertices in $A$:*

$$pr_v\big|_A \ge \sum_{t=0}^{\infty} \alpha(1-\alpha)^t \left(\chi_v - l\right) \widetilde{W}^t = \widetilde{pr}_v - \widetilde{pr}_l \ .$$

*Proof.* We begin the proof with a simple observation. The following inequality on vector holds coordinate-wisely on all vertices in $A$ according to the definition of $\widehat{W}$:

$$pr_v\big|_A = \sum_{t=0}^{\infty} \alpha(1-\alpha)^t \big(\chi_v W^t\big)\big|_A \ge \sum_{t=0}^{\infty} \alpha(1-\alpha)^t \chi_v \widehat{W}^t \ .$$

Therefore, to lower bound $pr_v\big|_A$ it suffices to lower bound the right hand side. Now owing to Claim 3.5 we further reduce the computation on matrix $\widehat{W}$ to that on matrix $\widetilde{W}$:

$$\chi_v \widehat{W}^t = \left(\chi_v \widehat{W}^{t-1} - l_t\right) \widetilde{W} = \left(\left(\chi_v \widehat{W}^{t-2} - l_{t-1}\right) \widetilde{W} - l_t\right) \widetilde{W} = \ldots = \chi_v \widetilde{W}^t - \sum_{j=1}^{t} l_j \widetilde{W}^{t-j+1} \ .$$

We next combine the above two inequalities and compute

$$\begin{aligned} pr_v\big|_A \ge \sum_{t=0}^{\infty} \alpha(1-\alpha)^t \chi_v \widehat{W}^t &= \sum_{t=0}^{\infty} \alpha(1-\alpha)^t \left(\chi_v \widetilde{W}^t - \sum_{j=1}^{t} l_j \widetilde{W}^{t-j+1}\right) \\ &= \sum_{t=0}^{\infty} \alpha(1-\alpha)^t \chi_v \widetilde{W}^t - \sum_{t=0}^{\infty} \alpha(1-\alpha)^t \sum_{j=1}^{t} l_j \widetilde{W}^{t-j+1} \\ &= \sum_{t=0}^{\infty} \alpha(1-\alpha)^t \chi_v \widetilde{W}^t - \sum_{j=1}^{\infty} (1-\alpha)^{j-1} l_j \sum_{t=1}^{\infty} \alpha(1-\alpha)^t \widetilde{W}^t \\ &\ge \sum_{t=0}^{\infty} \alpha(1-\alpha)^t \chi_v \widetilde{W}^t - \sum_{j=1}^{\infty} (1-\alpha)^{j-1} l_j \sum_{t=0}^{\infty} \alpha(1-\alpha)^t \widetilde{W}^t \\ &= \sum_{t=0}^{\infty} \alpha(1-\alpha)^t \left(\chi_v - \sum_{j=1}^{\infty}(1-\alpha)^{j-1} l_j\right) \widetilde{W}^t = \sum_{t=0}^{\infty} \alpha(1-\alpha)^t \left(\chi_v - l\right) \widetilde{W}^t \ . \end{aligned}$$

At last, we upper bound the one norm of $l$ using Claim 3.5 again:

$$\|l\|_1 = \sum_{j=1}^{\infty}(1-\alpha)^{j-1}\|l_j\|_1 = \sum_{j=1}^{\infty}(1-\alpha)^{j-1}(\texttt{Leak}(v,j) - \texttt{Leak}(v,j-1))$$
$$= \sum_{j=1}^{\infty}\alpha(1-\alpha)^{j-1}\texttt{Leak}(v,j) \leq \frac{2\phi(A)(1-\alpha)}{\alpha(1-\alpha)} = \frac{2\phi(A)}{\alpha} \ ,$$

where the last inequality uses (3.4). □

So far we have also shown (3.2) and this ends the proof of Lemma 3.1. ■

## 3.4 Proof of Lemma 3.2

**Lemma 3.2** (restated)**.** *When the teleport probability* $\alpha \leq \frac{\phi_{\mathsf{s}}(A)^2}{72(3+\log \mathrm{vol}(A))}$ *(or more weakly when* $\alpha \leq \frac{\lambda(A)}{9(3+\log \mathrm{vol}(A))}$*, or* $\alpha \leq O\big(\frac{1}{\tau_{\mathrm{mix}}}\big)$*), we have that*

$$\forall u \in A, \quad \widetilde{pr}_v(u) = \sum_{t=0}^{\infty}\alpha(1-\alpha)^t\left(\chi_v \widetilde{W}^t\right)(u) > \frac{4}{5}\frac{\deg_A(u)}{\mathrm{vol}(A)} \ .$$

*Proof.* We first prove this lemma in the case when $\alpha \leq \frac{\phi_{\mathsf{s}}(A)^2}{72(3+\log \mathrm{vol}(A))}$ or $\alpha \leq \frac{\lambda(A)}{9(3+\log \mathrm{vol}(A))}$. We will then extend it to the weakest assumption $\alpha \leq O\big(\frac{1}{\tau_{\mathrm{mix}}}\big)$. For a discussion on the comparisons between those three assumptions, see Section 2.1.

Recall that we defined $\widetilde{W}$ to be the lazy random walk matrix on $A$ with outgoing edges removed, and denoted by $\lambda = \lambda(A)$ the spectral gap on the lazy random walk matrix of $G[A]$ (cf. Section 2.1). Then, by the theory of infinity-norm mixing time of a Markov chain, the length-$t$ random walk starting at any vertex $v \in A$ will land in a vertex $u \in A$ with probability:

$$(\chi_v\widetilde{W}^t)(u) \geq \frac{\deg_A(u)}{\sum_{w\in A}\deg_A(w)} - (1-\lambda)^t\sqrt{\frac{\deg_A(v)}{\min_y \deg_A(y)}}$$
$$\geq \frac{\deg_A(u)}{\sum_{w\in A}\deg_A(w)} - (1-\lambda)^t\deg_A(v) \ .^7$$

Now if we choose $T_0 = \frac{3+\log \mathrm{vol}(A)}{\lambda}$ then for any $t \geq T_0$:

$$(\chi_v\widetilde{W}^t)(u) \geq \frac{9}{10}\frac{\deg_A(u)}{\sum_{w\in V}\deg_A(w)} \geq \frac{9}{10}\frac{\deg_A(u)}{\mathrm{vol}(A)} \ . \tag{3.5}$$

We then convert this into the language of PageRank vectors:

$$\sum_{t=0}^{\infty}\alpha(1-\alpha)^t(\chi_v\widetilde{W}^t)(u) \geq (1-\alpha)^{T_0}\alpha\sum_{t=0}^{\infty}(1-\alpha)^t(\chi_v\widetilde{W}^{t+T_0})(u)$$
$$\geq (1-\alpha)^{T_0}\alpha\sum_{t=0}^{\infty}(1-\alpha)^t\left(\frac{9}{10}\frac{\deg_A(u)}{\mathrm{vol}(A)}\right) = (1-\alpha)^{T_0}\left(\frac{9}{10}\frac{\deg_A(u)}{\mathrm{vol}(A)}\right) \ .$$

[7] Here we have used the fact that $\min_y \deg_A(y) \geq 1$. This is because otherwise $G[A]$ will be disconnected so that $\phi_{\mathsf{s}}(A) = 0, \lambda(A) = 0$ and $\tau_{\mathrm{mix}}(A) = \infty$, but none of the three can happen under our gap assumption $\mathsf{Gap}(A) \geq \Omega(1)$.

At last, we notice that $\alpha \le \frac{1}{9T_0}$ holds: this is either because we have chosen $\alpha \le \frac{\lambda(A)}{9(3+\log \mathrm{vol}(A))}$, or because we have chosen $\alpha \le \frac{\phi_{\mathsf{s}}(A)^2}{72(3+\log \mathrm{vol}(A))}$ and Cheeger's inequality $\lambda \ge \phi_{\mathsf{s}}(A)^2/8$ holds. As a consequence, it satisfies that $(1-\alpha)^{T_0} \ge 1 - \alpha T_0 \ge \frac{8}{9}$ and thus $(1-\alpha)^{T_0}\left(\frac{9}{10}\frac{\deg_A(u)}{\mathrm{vol}(A)}\right) \ge \frac{4}{5}\frac{\deg_A(u)}{\mathrm{vol}(A)}$.

We can also show our lemma under the assumption that $\alpha \le O(1/\tau_{\mathrm{mix}})$. In such a case, one can choose $T_0 = \Theta(\tau_{\mathrm{mix}})$ so that (3.5) and the rest of the proof still hold. It is worth emphasizing that since we always have $\frac{\phi_{\mathsf{s}}(A)^2}{\log \mathrm{vol}(A)} \le O\big(\frac{\lambda(A)}{\log \mathrm{vol}(A)}\big) \le O\big(\frac{1}{\tau_{\mathrm{mix}}}\big)$, this last assumption is the weakest one among all three. □

# 4 Guarantee Better Conductance

In the classical work of [ACL06], they have shown that when $\alpha = \Theta(\phi(A))$, among all sweep cuts on vector $p$ there exists one with conductance $O(\sqrt{\phi(A)\log n})$. In this section, we improve this result under our gap assumption $\mathsf{Gap}(A) \ge \Omega(1)$.

**Lemma 4.1.** *Letting $\alpha = \Theta(\mathsf{Conn}(A))$, among all sweep sets $S_c = \big\{u \in V : p(u) \ge c\frac{\deg(u)}{\mathrm{vol}(A)}\big\}$ for $c \in [\frac{1}{8}, \frac{1}{4}]$, there exists one, denoted by $S_{c^*}$, with conductance $\phi(S_{c^*}) = O(\phi(A)/\sqrt{\mathsf{Conn}(A)})$.*

*Proof sketch.* To convey the idea of the proof, we only consider the case when $p = pr_v$ is the exact PageRank vector, and the proof for the approximate case is a bit more involved and deferred to Section 4.1.

Let $E_0$ be the maximum value such that all sweep sets $S_c$ for $c \in [\frac{1}{8}, \frac{1}{4}]$ satisfy $|E(S_c, V \setminus S_c)| \ge E_0$, then it suffices to prove $E_0 \le O\big(\frac{\phi(A)}{\sqrt{\alpha}}\big)\mathrm{vol}(A)$. This is because, if so, then there exists some $S_{c^*}$ with $|E(S_{c^*}, V \setminus S_{c^*})| \le E_0$ and this combined with the result in Lemma 3.4 (i.e., $\mathrm{vol}(S_{c^*}) = (1 \pm O(1/\mathsf{Gap}(A)))\mathrm{vol}(A)$) gives

$$\phi(S_{c^*}) \le O\Big(\frac{E_0}{\mathrm{vol}(S_{c^*})}\Big) = O(\phi(A)/\sqrt{\alpha}) = O(\phi(A)/\sqrt{\mathsf{Conn}(A)}) \ .$$

We introduce some classical notations before we proceed in the proof. For any vector $q$ we denote by $q(S) \stackrel{\mathrm{def}}{=} \sum_{u\in S} q(u)$. Also, given a directed edge[8], $e = (a,b) \in E$ we let $p(e) = p(a,b) \stackrel{\mathrm{def}}{=} \frac{p(a)}{\deg(a)}$, and for a set of directed edges $E'$ we let $p(E') \stackrel{\mathrm{def}}{=} \sum_{e\in E'} p(e)$. We also let $E(A,B) \stackrel{\mathrm{def}}{=} \{(a,b) \in E \mid a \in A \wedge b \in B\}$ be the set of directed edges from $A$ to $B$.

Now for any set $S_{1/4} \subseteq S \subseteq S_{1/8}$, we compute that

$$\begin{aligned} p(S) = pr_v(S) &= \alpha\chi_v(S) + (1-\alpha)(pW)(S) \\ &\le \alpha + (1-\alpha)(pW)(S) \\ \implies (1-\alpha)p(S) &\le \alpha(1-p(S)) + (1-\alpha)(pW)(S) \\ \implies (1-\alpha)p(S) &\le 2\phi(A) + (1-\alpha)(pW)(S) \\ \implies p(S) &< O(\phi(A)) + (pW)(S) \ . \end{aligned} \tag{4.1}$$

Here we have used the fact that when $p = pr_v$ is exact, it satisfies $1 - p(S) = p(V - S) \le 2\phi(A)/\alpha$ according to Corollary 3.3. In the next step, we use the definition of the lazy random walk matrix

[8] $G$ is an undirected graph, but we study undirected edges with specific directions for analysis purpose only.

$W$ to compute that

$$
\begin{aligned}
(pW)(S) &= \Big( \sum_{(a,b)\in E(S,S)} p(a,b) + \sum_{(a,b)\in E(S,\bar{S})} \frac{p(a,b)+p(b,a)}{2} \Big) \\
&= \bigg( \frac{1}{2} p\Big(E(S,S)\Big) + \frac{1}{2} p\Big(E(S,S) \cup E(S,\bar{S}) \cup E(\bar{S},S)\Big) \bigg) \\
&\le \bigg( \frac{1}{2} p\Big[\big|E(S,S)\big|\Big] + \frac{1}{2} p\Big[\big|E(S,S) \cup E(S,\bar{S}) \cup E(\bar{S},S)\big|\Big] \bigg) \\
&= \bigg( \frac{1}{2} p\Big[\mathrm{vol}(S) - \big|E(S,\bar{S})\big|\Big] + \frac{1}{2} p\Big[\mathrm{vol}(S) + \big|E(\bar{S},S)\big|\Big] \bigg) \\
&\le \bigg( \frac{1}{2} p\big[\mathrm{vol}(S) - E_0\big] + \frac{1}{2} p\big[\mathrm{vol}(S) + E_0\big] \bigg) \quad . \qquad (4.2)
\end{aligned}
$$

Here the first inequality is due to the definition of the Lovász-Simonovits curve $p[x]$, and the second inequality is because $p[x]$ is concave. Next, suppose that in addition to $S_{1/4} \subseteq S \subseteq S_{1/8}$, we also know that $S$ is a sweep set, i.e., $\forall a \in S, b \notin S$ we have $\frac{p(a)}{\deg(a)} \ge \frac{p(b)}{\deg(b)}$. This implies $p(S) = p[\mathrm{vol}(S)]$ and combining (4.1) and (4.2) we obtain that

$$
\big(p[\mathrm{vol}(S)] - p\big[\mathrm{vol}(S) - E_0\big]\big) \le O(\phi(A)) + \big(p\big[\mathrm{vol}(S) + E_0\big] - p[\mathrm{vol}(S)]\big) \quad .
$$

Since we can choose $S$ to be an arbitrary sweep set between $S_{1/4}$ and $S_{1/8}$, we have that the inequality $p[x] - p[x - E_0] \le O(\phi(A)) + p[x + E_0] - p[x]$ holds for all end points $x \in [\mathrm{vol}(S_{1/4}), \mathrm{vol}(S_{1/8})]$ on the piecewise linear curve $p[x]$. This implies that the same inequality holds for any real number $x \in [\mathrm{vol}(S_{1/4}), \mathrm{vol}(S_{1/8})]$ as well. We are now ready to draw our conclusion by repeatedly applying this inequality. Letting $x_1 := \mathrm{vol}(S_{1/4})$ and $x_2 := \mathrm{vol}(S_{1/8})$, we have

$$
\begin{aligned}
\frac{E_0}{4\mathrm{vol}(A)} &\le p[x_1] - p[x_1 - E_0] \\
&\le O(\phi(A)) + (p[x_1 + E_0] - p[x_1]) \\
&\le 2 \cdot O(\phi(A)) + (p[x_1 + 2E_0] - p[x_1 + E_0]) \le \cdots \\
&\le \Big\lfloor \frac{x_2 - x_1}{E_0} + 1 \Big\rfloor O(\phi(A)) + (p[x_2 + E_0] - p[x_2]) \\
&\le \frac{\mathrm{vol}(S_{1/8} \setminus S_{1/4})}{E_0} O(\phi(A)) + \frac{E_0}{8\mathrm{vol}(A)} \\
&\le \frac{\mathrm{vol}(S_{1/8} \setminus A) + \mathrm{vol}(A \setminus S_{1/4})}{E_0} O(\phi(A)) + \frac{E_0}{8\mathrm{vol}(A)} \\
&\le \frac{O(\phi(A)/\alpha) \cdot \mathrm{vol}(A)}{E_0} O(\phi(A)) + \frac{E_0}{8\mathrm{vol}(A)} \quad ,
\end{aligned}
$$

where the first inequality uses the definition of $S_{1/4}$, the fifth inequality uses the definition of $S_{1/8}$, and last inequality uses Lemma 3.4 again. After re-arranging the above inequality we conclude that $E_0 \le O\big(\frac{\phi(A)}{\sqrt{\alpha}}\big)\mathrm{vol}(A)$ and finish the proof. □

The lemma above essentially shows the third property of Theorem 1 and finishes the proof of Theorem 1. For completeness of the paper, we still provide the formal proof for Theorem 1 below, and summarize our final algorithm in Algorithm 1. We are ready to put together all previous lemmas to show the main theorem of this paper.

**Algorithm 1** `PageRank-Nibble`

**Input:** $v, \mathsf{Conn}(A)$ and $\mathrm{vol}_0 \in [\frac{\mathrm{vol}(A)}{2}, \mathrm{vol}(A)]$.
**Output:** set $S$.

$\alpha \leftarrow \Theta(\mathsf{Conn}(A)) = \Theta(\phi(A) \cdot \mathsf{Gap}(A))$.
$p \leftarrow$ a $\frac{1}{10 \cdot \mathrm{vol}_0}$-approximate PageRank vector with starting vertex $v$ and teleport probability $\alpha$.
Sort all vertices in $\mathrm{supp}(p)$ according to $\frac{p(u)}{\deg(u)}$.
Consider all sweep sets $S'_c \stackrel{\mathrm{def}}{=} \{u \in \mathrm{supp}(p) : p(u) \geq \frac{c \deg(u)}{\mathrm{vol}_0}\}$ for $c \in [\frac{1}{8}, \frac{1}{2}]$, and let $S$ be the one among them with the best $\phi(S)$.

*Proof of Theorem 1.* As in Algorithm 1, we choose $\alpha = \Theta(\mathsf{Conn}(A))$ to satisfy the requirements of all previous lemmas. We define $A^g$ according to Lemma 3.1 and compute an $\varepsilon$-approximate PageRank vector starting from $v$ where $\varepsilon = \frac{1}{10\mathrm{vol}_0}$ satisfies (3.3).

Next we study all sweep sets $S'_c \stackrel{\mathrm{def}}{=} \{u \in \mathrm{supp}(p) : p(u) \geq \frac{c \deg(u)}{\mathrm{vol}_0}\}$ for $c \in [\frac{1}{16}, \frac{1}{4}]$. Notice that since $\mathrm{vol}_0 \in \big[\frac{\mathrm{vol}(A)}{2}, \mathrm{vol}(A)\big]$, all such sweep sets correspond to $S_d = \{u \in \mathrm{supp}(p) : p(u) \geq \frac{d \deg(u)}{\mathrm{vol}(A)}\}$ for some $d \in [\frac{1}{16}, \frac{1}{2}]$. Therefore, the output $S$ is also some $S_d$ sweep set with $d \in [\frac{1}{16}, \frac{1}{2}]$ and Lemma 3.4 guarantees the first two properties of the theorem.

On the other hand, Lemma 4.1 guarantees the existence of some sweep set $S_{d^*}$ satisfying $\phi(S_{d^*}) = O(\phi(A)/\sqrt{\mathsf{Conn}(A)})$. Since $d^* \in [\frac{1}{8}, \frac{1}{4}]$, this $S_{d^*}$ is also a sweep set $S'_c$ with $c \in [\frac{1}{16}, \frac{1}{4}]$, and must be considered as sweep set candidate in our Algorithm 1. This immediately implies that the output $S$ of Algorithm 1 must have a conductance $\phi(S)$ that is at least as good as $\phi(S_{d^*}) = O(\phi(A)/\sqrt{\mathsf{Conn}(A)})$, finishing the proof for the third property of the theorem.

At last, as a direct consequence of Proposition 2.3 and the fact that the computation of the approximate PageRank vector is the bottleneck for the running time, we conclude that Algorithm 1 runs in time $O(\frac{\mathrm{vol}(A)}{\alpha}) = O(\frac{\mathrm{vol}(A)}{\mathsf{Conn}(A)})$. □

## 4.1 Proof of Lemma 4.1

**Lemma 4.1.** *Letting $\alpha = \Theta(\mathsf{Conn}(A))$, among all sweep sets $S_c = \big\{u \in V : p(u) \geq c\frac{\deg(u)}{\mathrm{vol}(A)}\big\}$ for $c \in [\frac{1}{8}, \frac{1}{4}]$, there exists one, denoted by $S_{c^*}$, with conductance $\phi(S_{c^*}) = O(\phi(A)/\sqrt{\mathsf{Conn}(A)})$.*

*Proof.* We only point out how to extend our proof in the exact case to the case when $p$ is an $\varepsilon$-approximate PageRank vector. For any set $S_{1/4} \subseteq S \subseteq S_{1/8}$, we compute that

$$\begin{aligned} p(S) = pr_{\chi_v - r}(S) &= \alpha(\chi_v - r)(S) + (1-\alpha)(pW)(S) \\ &= \alpha(\chi_v - r)(V) + \alpha r(V \setminus S) + (1-\alpha)(pW)(S) \\ &\leq \alpha(\chi_v - r)(V) + \alpha\left(r(V \setminus A) + r(A \setminus S)\right) + (1-\alpha)(pW)(S) \\ &= \alpha p(V) + \alpha\left(r(V \setminus A) + r(A \setminus S)\right) + (1-\alpha)(pW)(S) \end{aligned}$$

where in the last equality we have used $(\chi_v - r)(V) = p(V)$, owing to the fact that $p = (\chi_v - r)\sum_{t=0}^{\infty} \alpha(1-\alpha)^t W^t$, but $W$ is a random walk matrix that preserves the total probability mass.

We next notice that $r(V \setminus A) \le \frac{2\phi(A)}{\alpha}$ according to Corollary 3.3, as well as

$$\begin{aligned}
r(A \setminus S) &\le \varepsilon \mathrm{vol}(A \setminus S) && \text{(according to Definition 2.2)} \\
&\le \varepsilon \left( \frac{2\phi(A)}{\alpha(\frac{3}{5} - \frac{1}{4})} + 8\phi(A) \right) \mathrm{vol}(A) && \text{(according to Lemma 3.4 and } S \supseteq S_{1/4}) \\
&< \frac{7\phi(A)}{\alpha} \varepsilon \mathrm{vol}(A) && \text{(using } \alpha \le \tfrac{1}{9} \text{ from the our choice in Section 3.4)} \\
&\le \frac{0.7\phi(A)}{\alpha} \ . && \text{(using our choice of } \varepsilon \le \tfrac{1}{10}\mathrm{vol}(A) \text{ in Section 3.2)}
\end{aligned}$$

Therefore, we have

$$\begin{aligned}
p(S) &\le \alpha p(V) + \alpha \Big( \frac{2\phi(A)}{\alpha} + \frac{0.7\phi(A)}{\alpha} \Big) + (1-\alpha)(pW)(S) \\
&= \alpha p(V) + 2.7\phi(A) + (1-\alpha)(pW)(S) \\
\implies (1-\alpha)p(S) &\le \alpha \cdot p(V \setminus S) + 2.7\phi(A) + (1-\alpha)(pW)(S) \\
\implies (1-\alpha)p(S) &\le 4.7\phi(A) + (1-\alpha)(pW)(S) && \text{(using Corollary 3.3)} \\
\implies p(S) &\le 5.3\phi(A) + (pW)(S) && \text{(using } \alpha \le \tfrac{1}{9} \text{ again)}
\end{aligned}$$

In sum, we have arrived at the same conclusion as (4.1) in the case when $p$ is only approximate, and the rest of the proof follows in the same way as in the exact case. □

# 5 Tightness of Our Analysis

It is a natural question to ask under our newly introduced assumption $\mathsf{Gap}(A) \ge \Omega(1)$: is $O(\phi(A)/\sqrt{\mathsf{Conn}(A)})$ the best conductance we can obtain from a local algorithm? We show that this is true if one sticks to a sweep-cut algorithm using PageRank vectors.

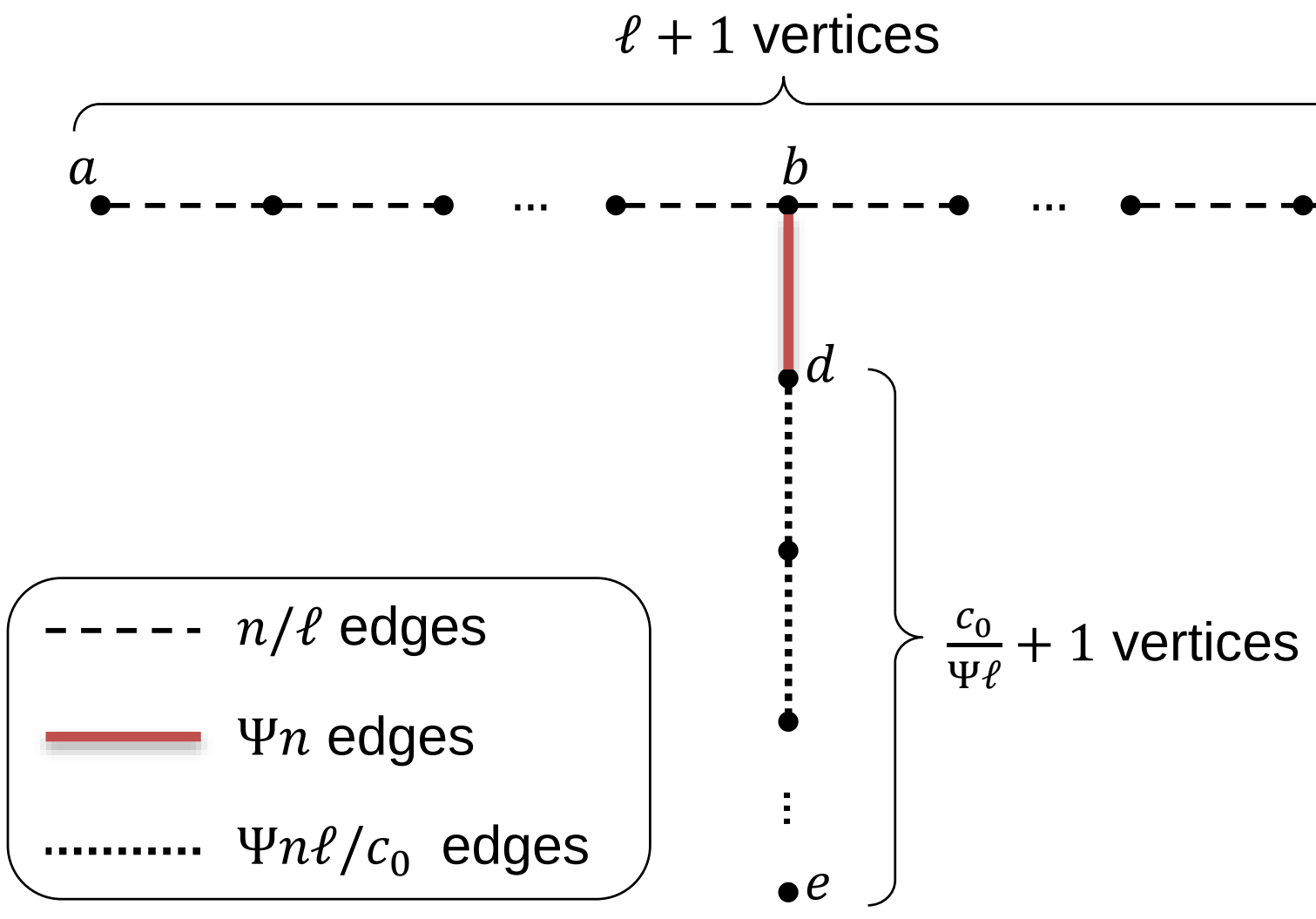

Figure 1: Our hard instance for proving tightness. One can pick for instance $\ell \approx n^{0.4}$ and $\phi(A) \approx \frac{1}{n^{0.9}}$, so that $n/\ell \approx n^{0.6}$, $\phi(A)n \approx n^{0.1}$ and $\phi(A)n\ell \approx n^{0.5}$.

More specifically, we show that our analysis in Section 4 is tight by constructing the following hard instance. Consider a (multi-)graph with two chains (see Figure 1) of vertices, and there are multi-edges connecting them.[9] In particular:

- the top chain (ended with vertex $a$ and $c$ and with midpoint $b$) consists of $\ell+1$ vertices where $\ell$ is even with $\frac{n}{\ell}$ edges between each consecutive pair;
- the bottom chain (ended with vertex $d$ and $e$) consists of $\frac{c_0}{\phi(A)\ell}+1$ vertices with $\frac{\phi(A)n\ell}{c_0}$ edges between each consecutive pair, where the constant $c_0$ is to be determined later; and
- vertex $b$ and $d$ are connected with $\phi(A)n$ edges.

We let the top chain to be our promised target cluster $A$. The total volume of $A$ is $2n+\phi(A)n$, while the total volume of the entire graph is $4n+2\phi(A)n$. The mixing time for $A$ is $\tau_{\text{mix}}(A)=\Theta(\ell^2)$, and the conductance $\phi(A)=\frac{\phi(A)n}{\text{vol}(A)}\approx\frac{\phi(A)}{2}$. Suppose that the gap assumption $\mathsf{Gap}(A)=\frac{1}{\tau_{\text{mix}}(A)\cdot\phi(A)}\approx\frac{1}{\phi(A)\ell^2}\gg 1$ is satisfied, i.e., $\phi(A)\ell^2=o(1)$. (For instance one can let $\ell\approx n^{0.4}$ and $\phi(A)\approx\frac{1}{n^{0.9}}$ to achieve this requirement.)

We then consider a PageRank random walk that starts at vertex $v=a$ and with teleport probability $\alpha=\frac{\gamma}{\ell^2}$ for some arbitrarily small constant $\gamma>0$.[10] Let $pr_a$ be this PageRank vector, and we prove in Appendix A the following lemma:

**Lemma 5.1.** *For any $\gamma\in(0,4]$ and letting $\alpha=\gamma/\ell^2$, there exists some constant $c_0$ such that when studying the PageRank vector $pr_a$ starting from vertex $a$ in Figure 1, the following holds* $\frac{pr_a(d)}{\deg(d)}>\frac{pr_a(c)}{\deg(c)}$.

This lemma implies that, for any sweep-cut algorithm based on this vector $pr_a$, even if it computes $pr_a$ exactly and looks for all possible sweep cuts, then none of them gives a better conductance than $O(\phi(A)/\sqrt{\mathsf{Conn}(A)})$. More specifically, for any sweep set $S$:

- if $c\notin S$, then $|E(S,V\setminus S)|$ is at least $\frac{n}{\ell}$ because it has to contain a (multi-)edge in the top chain. Therefore, the conductance $\phi(S)\geq\Omega(\frac{n}{\ell\text{vol}(S)})\geq\Omega(\frac{1}{\ell})\geq\Omega(\phi(A)/\sqrt{\mathsf{Conn}(A)})$; or
- if $c\in S$, then $d$ must be also in $S$ because it has a higher normalized probability than $c$ using Lemma 5.1. In this case, $|E(S,V\setminus S)|$ is at least $\frac{\phi(A)n\ell}{c_0}$ because it has to contain a (multi-)edge in the bottom chain. Therefore, the conductance $\phi(S)\geq\Omega(\frac{\phi(A)n\ell}{\text{vol}(S)})\geq\Omega(\phi(A)\ell)=\Omega(\phi(A)/\sqrt{\mathsf{Conn}(A)})$.

This ends the proof of Theorem 3. □

# 6 Empirical Evaluation

The PageRank local clustering method has been studied empirically in various previous work. For instance, Gleich and Seshadhri [GS12] performed experiments on 15 datasets and confirmed that PageRank outperformed many others in terms of conductance, including the famous `METIS` algorithm. Moreover, [LLDM09] studied PageRank against `METIS` + `MQI` which is the `METIS` algorithm plus a flow-based post-processing. Their experiments confirmed that although METIS + MQI outperforms PageRank in terms of conductance,[11] however, the PageRank algorithm's outputs are more "community-like", and they enjoy other desirable properties.

[9] One can transform this example into a graph without parallel edges by splitting vertices into expanders, but that goes out of the purpose of this section.

[10] Although we promised in Theorem 3 to study all starting vertices $v\in A$, in this version of the paper we only concentrate on $v=a$ because other choices of $v$ are only easier and can be analyzed similarly. In addition, this choice of $\alpha=\frac{\gamma}{\ell^2}$ is consistent with the one used Theorem 1.

Since our PageRank-Nibble is essentially the same PageRank method as before with only theoretical changes in the parameters, it certainly embraces the same empirical behavior as those literatures above. Therefore, in this section we perform experiments *only for the sake of* demonstrating our theoretical discoveries in Theorem 1, without comparisons to other methods. We run our algorithm against both synthetic and real datasets.

Recall that Theorem 1 has three properties. The first two properties are *accuracy guarantees* that ensure the output set $S$ well approximates $A$ in terms of volume; and the third property is a *cut-conductance guarantee* that ensures the output set $S$ has small $\phi(S)$. We now provide experimental results to support them.

**Experiment 1.** In the first experiment, we study a synthetic graph of 870 vertices. We carefully choose the parameters as follows in order to confuse the PageRank-Nibble algorithm so that it cannot identify $A$ up to a very high accuracy. We let the vertices be divided into three disjoint subsets: subset $A$ (which is the desired set) of 300 vertices, subset $B$ of 20 vertices and subset $C$ of 550 vertices. We assume that $A$ is constructed from the Watts-Strogatz model[12] with mean degree $K = 60$ and a parameter $\beta \in [0, 1]$ to control the connectivity of $G[A]$: varying $\beta$ makes it possible to interpolate between a regular lattice ($\beta = 0$) that is not-well-connected and a random graph ($\beta = 1$) that is well-connected. We then construct the rest of the graph by throwing in random edges, or more specifically, we add an edge

- with probability 0.3 between each pair of vertices in $B$ and $B$;
- with probability 0.02 between each pair of vertices in $C$ and $C$;
- with probability 0.001 between each pair of vertices in $A$ and $B$;
- with probability 0.002 between each pair of vertices in $A$ and $C$; and
- with probability 0.002 between each pair of vertices in $B$ and $C$.

It is not hard to verify that in this randomly generated graph, the (expected) conductance $\phi(A) = \phi(A)$ is independent of $\beta$. As a result, the larger $\beta$ is, we should expect the larger the well-connectedness $A$ enjoys, and therefore the larger $\mathsf{Gap}(A)$ is in Theorem 1. This should lead to a better performance both in terms of accuracy and conductance when $\beta$ goes larger.

To confirm this, we perform an experiment on this randomly generated graph with various choices of $\beta$. For each choice of $\beta$, we run our PageRank-Nibble algorithm with teleport probability $\alpha$ chosen to be the best one in the range of $[0.001, 0.3]$, starting vertex $v$ chosen to be a random one in $A$, and $\varepsilon$ to be sufficiently small. We then run our algorithm 100 times each time against a different random graph instance. We then plot in Figure 2 two curves (along with their 94% confidence intervals) as a function of $\beta$: the average conductance over $\phi(A)$ ratio, i.e., $\frac{\phi(S)}{\phi(A)}$, and the average clustering accuracy, i.e., $1 - \frac{|A\Delta S|}{|V|}$. Our experiment confirms our result in Theorem 1: PageRank-Nibble performs better both in accuracy and conductance as $\mathsf{Gap}(A)$ goes larger.

**Experiment 2.** In the second experiment, we use the USPS zipcode data set[13] that was also used in the work from [WLS+12]. Following their experiment, we construct a weighted $k$-NN graph with $k = 20$ out of this data set. The similarity between vertex $i$ and $j$ is computed as $w_{ij} = \exp(-d_{ij}^2/\sigma)$ if $i$ is within $j$'s $k$ nearest neighbors or vice versa, and $w_{ij} = 0$ otherwise, where $\sigma = 0.2 \times r$ and $r$ denotes the average square distance between each point to its 20th nearest neighbor.

[11] This is because MQI is designed to specifically shoot for conductance minimization using flow operations, see [LR04]. It is generalized by Andersen and Lang [AL08] and then made local by Orecchia and Zhu [OZ14].

[12] See http://en.wikipedia.org/wiki/Watts_and_Strogatz_model.

[13] http://www-stat.stanford.edu/~tibs/ElemStatLearn/data.html.

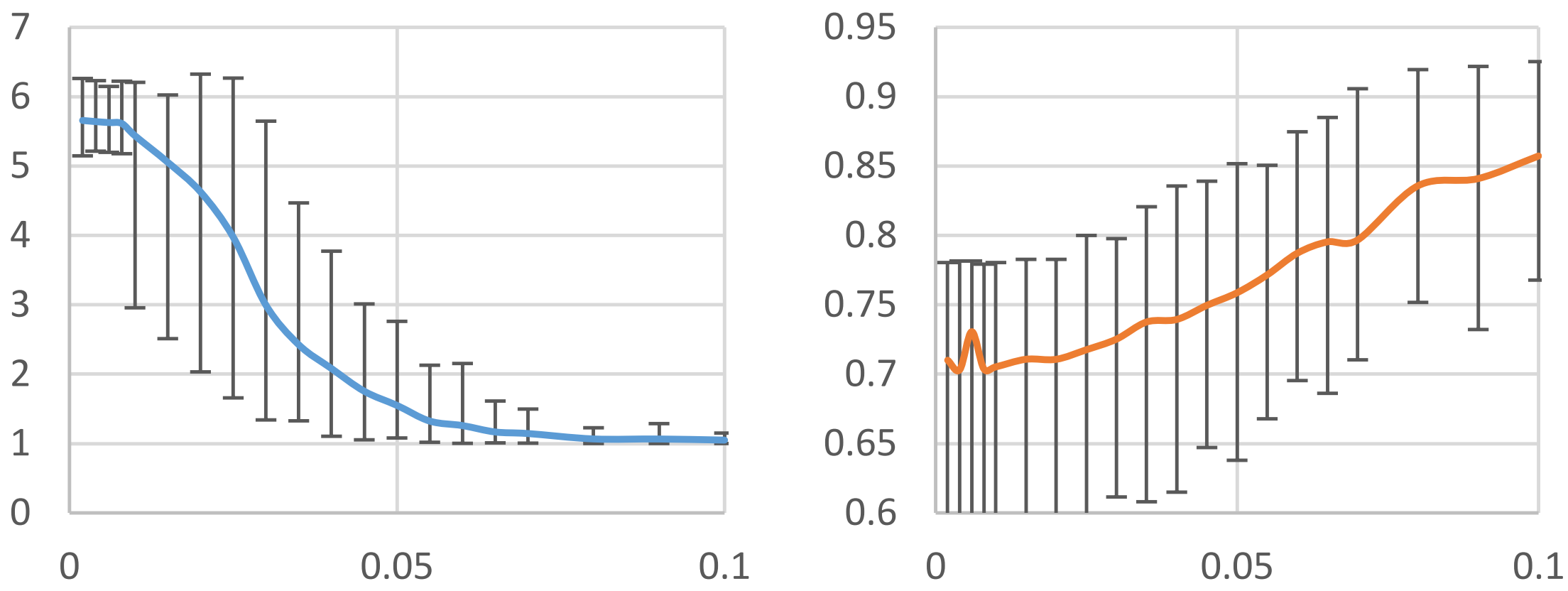

Figure 2: Experimental result on the synthetic data. The horizontal axis represents the value of $\beta$ for constructing our graph, the blue curve (left) represents the ratio $\frac{\phi(S)}{\phi(A)}$, and the red curve (right) represents the clustering accuracy. The vertical bars are 94% confidence intervals for 100 runs.

This is a dataset with 9298 images of handwritten digits between 0 to 9, and we treat it as 10 separate binary-classification problems. For each of them, we pick an arbitrary starting vertex in it, let $\alpha = 0.003$ and $\varepsilon = 0.00005$, and then run our `PageRank-Nibble` algorithm. We report our results in Table 1. For each of the 10 binary-classifications, we have a ground-truth set $A$ that contains all data points associated with the given digit. We then compare the conductance of our output set $\phi(S)$ against the desired conductance $\phi(A) = \phi(A)$, and our algorithm consistently outperforms the desired one on all 10 clusters. (Notice that it is possible to see an output set $S$ to have smaller conductance than $A$, because $A$ is not necessarily the sparest cut in the graph.) In addition, one can also confirm from Table 1 that our algorithm enjoys high precision and recall.

| Digit | 0 | 1 | 2 | 3 | 4 | 5 | 6 | 7 | 8 | 9 |
|---|---|---|---|---|---|---|---|---|---|---|
| $\phi(A) = \phi(A)$ | 0.00294 | 0.00304 | 0.08518 | 0.03316 | 0.22536 | 0.08580 | 0.01153 | 0.03258 | 0.09761 | 0.05139 |
| $\phi(S)$ | 0.00272 | 0.00067 | 0.03617 | 0.02220 | 0.00443 | 0.01351 | 0.00276 | 0.00456 | 0.03849 | 0.00448 |
| Precision | 0.993 | 0.995 | 0.839 | 0.993 | 0.988 | 0.933 | 0.946 | 0.985 | 0.941 | 0.994 |
| Recall | 0.988 | 0.988 | 0.995 | 0.773 | 0.732 | 0.896 | 0.997 | 0.805 | 0.819 | 0.705 |

Table 1: Clustering results on the USPS zipcode data set. We report precision $|A \cap S|/|S|$ and recall $|A \cap S|/|A|$.

## Acknowledgements

We thank Lorenzo Orecchia, Jon Kelner, Aditya Bhaskara for helpful conversations. This work is partly supported by Google and a Simons award (grant no. 284059).

# Appendix

## A Missing Proofs in Section 5

In this section we show that our conductance analysis for Theorem 1 is tight. We emphasize here that such a tightness proof is very non-trivial, because one has to provide a graph hard instance and start to upper and lower bound the probabilities of reaching specific vertices up to a very high precision. This is different from the mixing time theory on Markov chains, as for instance, on a chain of $\ell$ vertices it is known that a random walk of $O(\ell^2)$ steps mixes, but in addition we need to compute how faster it mixes on one vertex than another vertex.

In Appendix A.1 we begin with some warm-up lemmas for the PageRank vector on a single chain, and then in Appendix A.2 we formally prove Lemma 5.1 with the help from those lemmas.

### A.1 Useful Lemmas for a PageRank Random Walk on a Chain

In this subsection we provide four useful lemmas about a PageRank random walk on a single chain. For instance, in the first of them we study a chain of length $\ell$ and compute an upper bound on the probability to reach the rightmost vertex from the leftmost one. The other three lemmas are similar in this format. Those lemmas require the study of the eigensystem of a lazy random walk matrix on this chain, followed by very careful but problem-specific analyses.

**Lemma A.1.** *Let $\ell$ be an even integer, and consider a chain of $\ell+1$ vertices with the leftmost vertex indexed by $0$ and the rightmost vertex indexed by $\ell$. Let $pr_{\chi_0}$ be the PageRank vector for a random walk starting at vertex $0$ with teleport probability $\alpha = \frac{\gamma}{\ell^2}$ for some constant $\gamma$. Then,*

$$pr_{\chi_0}(\ell) \le \frac{1}{2\ell}\left(1 - \frac{2\gamma}{\pi^2/4+\gamma} + \frac{2\gamma}{\pi^2+\gamma} + O\Big(\frac{1}{\ell^2}\Big)\right) .$$

*Proof.* Let us define

$$W = \begin{pmatrix} \frac{1}{2} & \frac{1}{2} & & & \\ \frac{1}{4} & \frac{1}{2} & \frac{1}{4} & & \\ & \frac{1}{4} & \frac{1}{2} & \ddots & \\ & & \ddots & \ddots & \frac{1}{4} \\ & & & \frac{1}{2} & \frac{1}{2} \end{pmatrix}$$

to be the $(\ell+1)\times(\ell+1)$ lazy random walk matrix of our chain. For $k = 0, 1, \dots, \ell$, define:

$$\lambda_k \stackrel{\text{def}}{=} \frac{1+\cos(\frac{\pi k}{\ell})}{2} = \cos^2\Big(\frac{\pi k}{2\ell}\Big)$$
$$v_k(u) \stackrel{\text{def}}{=} \deg(u)\cdot\cos\Big(\frac{\pi k u}{\ell}\Big) \qquad (u = 0, 1, \dots, \ell) , \tag{A.1}$$

where $\deg(u)$ is the degree for the $u$-th vertex, that is, $\deg(0) = \deg(\ell) = 1$ while $\deg(i) = 2$ for $i \in \{1, 2, \dots, \ell-1\}$. Then it is routinary to verify that $v_k \cdot W = \lambda_k \cdot v_k$ and thus

*$v_k$ is the $k$-th (left-)eigenvector and $\lambda_k$ is the $k$-th eigenvalue for matrix $W$.*

We remark here that since $W$ is not symmetric, those eigenvectors are not orthogonal to each other in the standard basis. However, under the notion of inner product $\langle x, y\rangle \stackrel{\text{def}}{=} \sum_{i=0}^{\ell} x(i)y(i)\deg(i)^{-1}$, they form an orthonormal basis.

It now expand our starting probability vector $\chi_0$ under this orthonormal basis:

$$\chi_0 = (1, 0, 0, \dots, 0) = \frac{1}{2\ell}\left(v_0 + 2\sum_{k=1}^{\ell-1} v_k + v_\ell\right) \quad .$$

As a consequence when $t > 0$, using $\lambda_\ell = 0$:

$$\chi_0 W^t = \frac{1}{2\ell}\left(v_0 + 2\sum_{k=1}^{\ell-1}(\lambda_k)^t v_k\right) \quad .$$

Now it is easy to compute the exact probability of reaching the right-most vertex $\ell$:

$$\begin{aligned}
\chi_0 W^t(\ell) &= \frac{1}{2\ell}\left(v_0(\ell) + 2\sum_{k=1}^{\ell-1}(\lambda_k)^t v_k(\ell)\right) = \frac{1}{2\ell}\left(1 + 2\sum_{k=1}^{\ell-1}\cos^{2t}\left(\frac{\pi k}{2\ell}\right)\cos(\pi k)\right) \\
&= \frac{1}{2\ell}\left(1 + 2\sum_{k=1}^{\ell-1}\cos^{2t}\left(\frac{\pi k}{2\ell}\right)(-1)^k\right) \le \frac{1}{2\ell}\left(1 - 2\cos^{2t}(\frac{\pi}{2\ell}) + 2\cos^{2t}\left(\frac{\pi}{\ell}\right)\right) \quad .
\end{aligned}$$

At last, we translate this language into the PageRank vector $pr_{\chi_0}$ and obtain

$$\begin{aligned}
pr_{\chi_0}(\ell) = \sum_{t=0}^{\infty}\alpha(1-\alpha)^t\chi_0 W^t(\ell) &\le \frac{1}{2\ell}\left(\alpha v_\ell(\ell) + \sum_{t=0}^{\infty}\alpha(1-\alpha)^t\left(1 - 2\cos^{2t}\left(\frac{\pi}{2\ell}\right) + 2\cos^{2t}\left(\frac{\pi}{\ell}\right)\right)\right) \\
&= \frac{1}{2\ell}\left(\alpha + 1 - \frac{2\alpha}{1-(1-\alpha)\cos^2(\frac{\pi}{2\ell})} + \frac{2\alpha}{1-(1-\alpha)\cos^2(\frac{\pi}{\ell})}\right) \\
&\le \frac{1}{2\ell}\left(1 - \frac{2\gamma}{\pi^2/4+\gamma} + \frac{2\gamma}{\pi^2+\gamma} + O\Big(\frac{1}{\ell^2}\Big)\right) \quad .
\end{aligned}$$

We remark here that the last inequality is obtained using Taylor approximation. □

**Lemma A.2.** *Let $\ell$ be an even integer, and consider a chain of $\ell+1$ vertices with the leftmost vertex indexed by $0$ and the rightmost vertex indexed by $\ell$. Let $pr_{\chi_0}$ be the PageRank vector for a random walk starting at vertex $0$ with teleport probability $\alpha = \frac{\gamma}{\ell^2}$ for some constant $\gamma$. Then,*

$$pr_{\chi_0}\Big(\frac{\ell}{2}\Big) \ge \frac{1}{\ell}\left(1 - \frac{2\gamma}{\pi^2+\gamma} - O\Big(\frac{1}{\ell^2}\Big)\right) \quad .$$

*Proof.* Recall from the proof of Lemma A.1 that for $t > 0$ we have

$$\chi_0 W^t = \frac{1}{2\ell}\left(v_0 + 2\sum_{k=1}^{\ell-1}(\lambda_k)^t v_k\right) \quad .$$

Now it is easy to compute the exact probability of reaching the middle vertex $\frac{\ell}{2}$:

$$\begin{aligned}
\chi_0 W^t\Big(\frac{\ell}{2}\Big) &= \frac{1}{2\ell}\left(v_0\Big(\frac{\ell}{2}\Big) + 2\sum_{k=1}^{\ell-1}(\lambda_k)^t v_k\Big(\frac{\ell}{2}\Big)\right) = \frac{1}{\ell}\left(1 + 2\sum_{k=1}^{\ell-1}\cos^{2t}\left(\frac{\pi k}{2\ell}\right)\cos\left(\frac{\pi k}{2}\right)\right) \\
&= \frac{1}{\ell}\left(1 + 2\sum_{q=1}^{\ell/2-1}\cos^{2t}\left(\frac{2\pi q}{2\ell}\right)(-1)^q\right) \ge \frac{1}{\ell}\left(1 - 2\cos^{2t}\left(\frac{\pi}{\ell}\right)\right) \quad .
\end{aligned}$$

At last, we translate this language into the PageRank vector $pr_{\chi_0}$ and obtain

$$pr_{\chi_0}\Big(\frac{\ell}{2}\Big) = \sum_{t=0}^{\infty}\alpha(1-\alpha)^t\chi_0 W^t\Big(\frac{\ell}{2}\Big) \geq \frac{1}{\ell}\left(\alpha v_\ell\Big(\frac{\ell}{2}\Big) + \sum_{t=0}^{\infty}\alpha(1-\alpha)^t\left(1-2\cos^{2t}\left(\frac{\pi}{\ell}\right)\right)\right)$$
$$= \frac{1}{\ell}\left(\alpha v_\ell\Big(\frac{\ell}{2}\Big) + 1 - \frac{2\alpha}{1-(1-\alpha)\cos^2\left(\frac{\pi}{\ell}\right)}\right)$$
$$\geq \frac{1}{\ell}\left(1 - \frac{2\gamma}{\pi^2+\gamma} - O\Big(\frac{1}{\ell^2}\Big)\right) \quad .$$

We remark here that the last inequality is obtained using Taylor approximation. □

**Lemma A.3.** *Let $\ell$ be an even integer, and consider a chain of $\ell+1$ vertices with the leftmost vertex indexed by $0$ and the rightmost vertex indexed by $\ell$. Let $pr_{\chi_{\ell/2}}$ be the PageRank vector for a random walk starting at the middle vertex $\ell/2$ with teleport probability $\alpha = \frac{\gamma}{\ell^2}$ for some constant $\gamma$. Then,*

$$pr_{\chi_{\ell/2}}\Big(\frac{\ell}{2}\Big) \leq \frac{1}{\ell}\left(1+\sqrt{\gamma}+O\Big(\frac{1}{\ell}\Big)\right) \quad .$$

*Proof.* Following the notion of $\lambda_k$ and $v_k$ in (A.1), we expand our starting probability vector $\chi_{\ell/2}$ under this orthonormal basis:

$$\chi_{\ell/2} = (0,\dots,0,1,0,\dots,0) = \frac{1}{2\ell}\left(v_0 + 2\sum_{q=1}^{\ell/2-1}(-1)^q v_{2q} + (-1)^{\ell/2}v_\ell\right) \quad .$$

Then similar to the proof of Lemma A.1 we have that for all $t>0$

$$\chi_{\ell/2}W^t = \frac{1}{2\ell}\left(v_0 + 2\sum_{q=1}^{\ell/2-1}(-1)^q(\lambda_{2q})^t v_{2q}\right) \quad .$$

Now it is easy to compute the exact probability of reaching the middle vertex $\frac{\ell}{2}$:

$$\chi_{\ell/2}W^t\Big(\frac{\ell}{2}\Big) = \frac{1}{2\ell}\left(v_0\Big(\frac{\ell}{2}\Big) + 2\sum_{q=1}^{\ell/2-1}(-1)^q(\lambda_{2q})^t v_{2q}\Big(\frac{\ell}{2}\Big)\right) = \frac{1}{\ell}\left(1+2\sum_{q=1}^{\ell/2-1}(-1)^q\cos^{2t}\left(\frac{2\pi q}{2\ell}\right)\cos\left(\frac{2\pi q}{2}\right)\right)$$
$$= \frac{1}{\ell}\left(1+2\sum_{q=1}^{\ell/2-1}\cos^{2t}\left(\frac{2\pi q}{2\ell}\right)\right) = \frac{1}{\ell}\left(\frac{\ell}{2^{2t}}\sum_{k=-\lfloor t/\ell\rfloor}^{\lfloor t/\ell\rfloor}\binom{2t}{t+k\ell}\right) \quad .$$

Notice that in the last equality we have used a recent result on power sum of cosines that can be found in Theorem 1 of [Mer12]. Next we perform some classical tricks on binomial coefficients:

$$\sum_{k=-\lfloor t/\ell\rfloor}^{\lfloor t/\ell\rfloor}\binom{2t}{t+k\ell} = \binom{2t}{t} + 2\sum_{k=1}^{\lfloor t/\ell\rfloor}\binom{2t}{t+k\ell}$$
$$\leq \binom{2t}{t} + 2\sum_{k=1}^{\lfloor t/\ell\rfloor}\frac{1}{\ell}\left(\binom{2t}{t+(k-1)\ell+1} + \binom{2t}{t+(k-1)\ell+2} + \cdots + \binom{2t}{t+k\ell}\right)$$
$$\leq \binom{2t}{t} + \frac{1}{\ell}\sum_{q=0}^{2t}\binom{2t}{q} \leq \frac{2^{2t}}{\sqrt{\pi t}} + \frac{2^{2t}}{\ell} \quad ,$$

and in the last inequality we have used a famous upper bound on the central binomial coefficient that says $\binom{2t}{t} \leq \frac{2^{2t}}{\sqrt{\pi t}}$ for any integer $t \geq 1$ and $p \in \{0, 1, \ldots, 2t\}$.

At last, we translate this language into the PageRank vector $pr_{\chi_{\ell/2}}$ and obtain

$$\begin{aligned}
pr_{\chi_{\ell/2}}\Big(\frac{\ell}{2}\Big) &= \sum_{t=0}^{\infty} \alpha(1-\alpha)^t \chi_{\ell/2} W^t \Big(\frac{\ell}{2}\Big) \leq \alpha + \frac{1}{\ell}\left(\sum_{t=1}^{\infty} \alpha(1-\alpha)^t \frac{\ell}{2^{2t}}\left(\frac{2^{2t}}{\sqrt{\pi t}} + \frac{2^{2t}}{\ell}\right)\right) \\
&= \alpha + \frac{1}{\ell}\left(1 + \sum_{t=1}^{\infty} \alpha(1-\alpha)^t \frac{\ell}{\sqrt{\pi t}}\right) \leq \alpha + \frac{1}{\ell}\left(1 + \int_{t=0}^{\infty} \alpha(1-\alpha)^t \frac{\ell}{\sqrt{\pi t}} dt\right) \\
&= \alpha + \frac{1}{\ell}\left(1 + \frac{\alpha \ell}{\sqrt{-\log(1-\alpha)}}\right) \leq \frac{1}{\ell}\left(1 + \sqrt{\gamma} + O\Big(\frac{1}{\ell}\Big)\right) \enspace .
\end{aligned}$$

We remark here that the last inequality is obtained using Taylor approximation. □

**Lemma A.4.** *Consider an* infinite *chain with one special vertex called the* origin. *Note that the chain is infinite both to the left and to the right of the origin. Now we study the PageRank random walk on this infinite chain that starts from the origin with teleport probability $\alpha = \frac{\gamma}{\ell^2}$, and denote by $pr_{\chi_0}(0)$ be the probability of reaching the origin. Then,*

$$pr_{\chi_0}(0) \geq \frac{\sqrt{\pi\gamma}}{2\ell} - O\Big(\frac{1}{\ell^2}\Big) \enspace .$$

*Proof.* As before we begin with the analysis of a lazy random walk of a fixed length $t$, and will translate it into the language of a PageRank random walk in the end. Suppose in the $t$ actual number of steps, there are $t_1 \leq t$ number of them in which the random walk moves either to the left or to the right, while in the remaining $t - t_1$ of them the random walk stays. This happens with probability $\binom{t}{t_1}2^{-t}$. When $t_1$ is fixed, to reach the origin it must be the case that among $t_1$ left-or-right moves, exactly $t_1/2$ of them are left moves, and the other half are right moves. This happens with probability $\binom{t_1}{t_1/2}2^{-t_1}$. In sum, the probability to reach the origin in a $t$-step lazy random walk is:

$$\sum_{t_1=0}^{t} \binom{t}{t_1} 2^{-t} \binom{t_1}{t_1/2} 2^{-t_1} = \sum_{y=0}^{t/2} \binom{2y}{y}\binom{t}{2y} 2^{-2y-t} = \frac{1}{(t)!}\frac{(2t-1)!!}{2^t} = \frac{1}{(t)!} \cdot \frac{(2t)!}{t! 2^{2t}} = \binom{2t}{t} 2^{-2t} \geq \frac{1}{\sqrt{4t}} \enspace .$$

Here in the last inequality we have used the famous lower bound on the central binomial coefficient that says $\binom{2t}{t} \geq \frac{2^{2t}}{\sqrt{4t}}$ for $t \geq 1$. At last, we translate this into the language of a PageRank random walk:

$$\begin{aligned}
pr_{\chi_0}(0) \geq \alpha + \sum_{t=1}^{\infty} \alpha(1-\alpha)^t \frac{1}{\sqrt{4t}} \geq \alpha + \int_{t=1}^{\infty} \alpha(1-\alpha)^t \frac{1}{\sqrt{4t}} dt = \alpha + \frac{\alpha\sqrt{\pi}\left(1 - \mathrm{erf}\big(\sqrt{-\log(1-\alpha)}\big)\right)}{2\sqrt{-\log(1-\alpha)}} \\
\geq \frac{\sqrt{\pi\gamma}}{2\ell} - O\Big(\frac{1}{\ell^2}\Big) \enspace .
\end{aligned}$$

Here in the last inequality we have used the Taylor approximation for the Gaussian error function erf. □

## A.2 Proof of Lemma 5.1

We are now ready to show the proof for Lemma 5.1.

**Lemma 5.1.** *For any $\gamma \in (0,4]$ and letting $\alpha = \gamma/\ell^2$, there exists some constant $c_0$ such that when studying the PageRank vector $pr_a$ starting from vertex $a$ in Figure 1, it satisfies that $\frac{pr_a(d)}{\deg(d)} > \frac{pr_a(c)}{\deg(c)}$.*

We divide the proof into four steps. In the first step we provide an upper bound on $\frac{pr_a(c)}{\deg(c)}$ for vertex $c$, and in the second step we provide a lower bound on $\frac{pr_a(b)}{\deg(b)}$ for vertex $b$. Both these steps require a careful study on a finite chain (and in fact the top chain in Figure 1) which we have already done in Appendix A.1. They together will imply that

$$\frac{pr_a(b)}{\deg(b)} > (1+\Omega(1))\frac{pr_a(c)}{\deg(c)} \ . \tag{A.2}$$

In the third step, we show that

$$\frac{pr_a(d)}{\deg(d)} > (1-O(1))\frac{pr_a(b)}{\deg(b)} \ , \tag{A.3}$$

that is, the (normalized) probability for reaching $d$ must be roughly as large as $b$. This is a result of the fact that, suppose towards contradiction that $\frac{pr_a(d)}{\deg(d)}$ is much smaller than $\frac{pr_a(b)}{\deg(b)}$, then there must be a large amount of probability mass moving from $b$ to $d$ due to the nature of PageRank random walk, while a large fraction of them should remain at vertex $d$ due to the chain at the bottom, giving a contradiction to $\frac{pr_a(d)}{\deg(d)}$ being small.

And in the last step, we choose the constants very carefully to deduce $\frac{pr_a(d)}{\deg(d)} > \frac{pr_a(c)}{\deg(c)}$ out of (A.2) and (A.3).

**Step 1: upper bounding $pr_a(c)/\deg(c)$.** In the first step we upper bound the probability of reaching vertex $c$. Since removing the edges between $b$ and $d$ will disconnect the graph and thus only increase such probability, it suffices for us to consider just the top chain, which is equivalent to the PageRank random walk on a finite chain of length $\ell+1$ studied in Lemma A.1. In our language, taking into account the multi-edges, we have that

$$\begin{aligned}\frac{pr_a(c)}{\deg(c)} &\leq \frac{1}{n/\ell}\frac{1}{2\ell}\left(1-\frac{2\gamma}{\pi^2/4+\gamma}+\frac{2\gamma}{\pi^2+\gamma}+O\Big(\frac{1}{\ell^2}\Big)\right) \\ &= \frac{1}{2n}\left(1-\frac{2\gamma}{\pi^2/4+\gamma}+\frac{2\gamma}{\pi^2+\gamma}+O\Big(\frac{1}{\ell^2}\Big)\right) \ .\end{aligned} \tag{A.4}$$

**Step 2: lower bounding $pr_a(b)/\deg(b)$.** In this step we ask for help from a variant of Lemma 3.1. Letting $\widetilde{pr}_s$ be the PageRank vector on the induced subgraph $G[A]$ starting from $s$ with teleport probability $\alpha$, then Lemma 3.1 (and its actual proof) implies that $pr_a(b) \geq \widetilde{pr}_a(b) - \widetilde{pr}_l(b)$ where $l$ is a vector that is only non-zero at the boundary vertex $b$, and in addition, $\|l\|_1 = l(b) \leq \frac{2\phi(A)}{\alpha}$ since $a \in A^g$ is a good starting vertex. We can rewrite this as

$$pr_a(b) \geq \widetilde{pr}_a(b) - \frac{2\phi(A)}{\alpha}\widetilde{pr}_b(b) \ .$$

Next we use Lemma A.2 and Lemma A.3 to deduce that:

$$pr_a(b) \geq \frac{1}{\ell}\left(1-\frac{2\gamma}{\pi^2+\gamma}-O\Big(\frac{1}{\ell^2}\Big)\right) - \frac{2\phi(A)}{\alpha}\frac{1}{\ell}\left(1+\sqrt{\gamma}+O\Big(\frac{1}{\ell}\Big)\right) \ .$$

At last, we normalize this probability by its degree $\deg(b) = 2n/\ell + \phi(A)n$ and get:

$$\begin{aligned}
\frac{pr_a(b)}{\deg(b)} &\geq \frac{1}{2n + \phi(A)n\ell}\left(1 - \frac{2\gamma}{\pi^2+\gamma} - O\Big(\frac{1}{\ell^2}\Big) - \frac{2\phi(A)}{\alpha}\left(1 + \sqrt{\gamma} + O\Big(\frac{1}{\ell}\Big)\right)\right) \\
&\geq \frac{1}{2n}\left(1 - \frac{2\gamma}{\pi^2+\gamma} - O(\phi(A)\ell^2)\right) \ . \qquad \text{(A.5)}
\end{aligned}$$

**Step 3: lower bounding $pr_a(d)/\deg(d)$.** Since we have already shown a good lower bound on $pr_a(b)/\deg(b)$ in the previous step, one may naturally guess that a similar lower bound should apply to vertex $d$ as well because $b$ and $d$ are neighbors. This is not true in general, for instance if $d$ were connected to a very large complete graph then all probability mass that reached $d$ would be badly diluted. However, with our careful choice of the bottom chain, we will show that this is true in our case.

**Lemma A.5.** *Let $p^* \stackrel{\text{def}}{=} \frac{pr_a(b)}{\deg(b)}$, then either $\frac{pr_a(d)}{\deg(d)} \geq (1-c_1)p^*$ or $\frac{pr_a(d)}{\deg(d)} \geq \frac{c_1 c_0}{2} p^*(1 - O(\frac{1}{\ell}))$.*

*Proof.* Throughout the proof we assume that $\frac{pr_a(d)}{\deg(d)} < (1-c_1)p^*$ because otherwise we are done. Therefore, we only need to show that $\frac{pr_a(d)}{\deg(d)} \geq \frac{c_1 c_0}{2} p^*(1 - O(\frac{1}{\ell}))$ is true under this assumption.

We first show a lower bound on the amount of *net* probability that will leak from $A$ during the given PageRank random walk, i.e., `NetLeakage` $\stackrel{\text{def}}{=} \sum_{u \notin A} pr_a(u)$. Loosely speaking, this net probability is the amount of probability that will leak from $A$, subtracted by the amount of probability that will come back to $A$.

We introduce some notation first. Let $p^{(t)} \stackrel{\text{def}}{=} \chi_a W^t$ be the lazy random walk vector after $t$ steps, and using the similar notation as Lemma 4.1, we let $p^{(t)}(b,d) \stackrel{\text{def}}{=} \frac{p^{(t)}(b)}{\deg(b)}$ be the amount of probability mass sent from $b$ to $d$ per edge at time step $t$ to $t+1$, and similarly $p^{(t)}(d,b) \stackrel{\text{def}}{=} \frac{p^{(t)}(d)}{\deg(b)}$. If the PageRank random walk runs for a total of $t$ steps (which happens with probability $\alpha(1-\alpha)^t$), then the total amount of net leakage becomes $\sum_{i=0}^{t-1}\big(p^{(i)}(b,d) - p^{(i)}(d,b)\big) \cdot \phi(A)n$. This gives another way to compute the total amount of net leakage of a PageRank random walk:

$$\begin{aligned}
\texttt{NetLeakage} &= \sum_{t=0}^{\infty} \alpha(1-\alpha)^t \sum_{i=0}^{t-1} \big(p^{(i)}(b,d) - p^{(i)}(d,b)\big) \cdot \phi(A)n \\
&= \sum_{i=0}^{\infty} \big(p^{(i)}(b,d) - p^{(i)}(d,b)\big) \cdot \phi(A)n \sum_{t=i+1}^{\infty} \alpha(1-\alpha)^t \\
&= \sum_{i=0}^{\infty} \big(p^{(i)}(b,d) - p^{(i)}(d,b)\big) \cdot \phi(A)n \cdot (1-\alpha)^{i+1} \\
&= \frac{1-\alpha}{\alpha} \sum_{i=0}^{\infty} \alpha(1-\alpha)^i \big(p^{(i)}(b,d) - p^{(i)}(d,b)\big) \cdot \phi(A)n \\
&= \frac{1-\alpha}{\alpha}\left(\frac{pr_a(b)}{\deg(b)} - \frac{pr_a(d)}{\deg(d)}\right) \cdot \phi(A)n \geq \frac{1-\alpha}{\alpha} c_1 p^* \phi(A)n \ . \qquad \text{(A.6)}
\end{aligned}$$

Now we have a decent lower bound on the amount of net leakage, and we want to further lower bound $pr_a(d)$ using this `NetLeakage` quantity. We achieve so by studying an auxiliary "random walk" procedure $q^{(t)}$, where $q^{(0)} = p^{(0)} = \chi_a$, but

$$q^{(t+1)} \stackrel{\text{def}}{=} q^{(t)}W + \delta^{(t)}, \text{ where } \delta^{(t)}(u) \begin{cases} 0, & \text{if } u \neq b, u \neq d; \\ p^{(t)}(d,b) \cdot \phi(A)n, & \text{if } u = b; \\ -p^{(t)}(b,d) \cdot \phi(A)n, & \text{if } u = d. \end{cases}$$

It is not hard to prove by induction that for all $t \geq 0$, it satisfies $q^{(t)}(u) = p^{(t)}(u)$ for $u \in A$ and $q^{(t)}(u) = 0$ for $u \notin A$.[14] Then we have that:

$$\Delta \stackrel{\text{def}}{=} \sum_{t=0}^{\infty} \alpha(1-\alpha)^t q^{(t)} - pr_a$$

is precisely the vector that is zero everywhere in $A$ and equal to $pr_a$ everywhere in $V \setminus A$. We further notice that

$$\begin{aligned}\Delta &= \sum_{t=0}^{\infty} \alpha(1-\alpha)^t \left(q^{(t)} - p^{(t)}\right) = \sum_{t=0}^{\infty} \alpha(1-\alpha)^t \left(\sum_{i=0}^{t-1} \delta^{(i)} W^{t-i-1}\right) \\ &= \sum_{k=0}^{\infty}\sum_{i=0}^{\infty} \alpha(1-\alpha)^{k+i+1}\delta^{(i)} W^k = \sum_{k=0}^{\infty} \alpha(1-\alpha)^k \left(\sum_{i=0}^{\infty}(1-\alpha)^{i+1}\delta^{(i)}\right) W^k \ .\end{aligned}$$

Therefore, as long as we define $\delta \stackrel{\text{def}}{=} \sum_{i=0}^{\infty}(1-\alpha)^{i+1}\delta^{(i)} = \frac{1-\alpha}{\alpha}\sum_{i=0}^{\infty}\alpha(1-\alpha)^i\delta^{(i)}$, we can write $\Delta = pr_\delta$ also as a PageRank vector. We highlight here that $\delta$ is a vector that is non-zero only at vertex $b$ and $d$ (and in fact $\delta(d) \geq 0$ and $\delta(b) \leq 0$), such that $\delta(d) + \delta(b) = \texttt{NetLeakage}$ according to the first equality in (A.6).

Now we are ready to lower bound $pr_a(d)$. Using the linearity of PageRank vectors we have

$$pr_a(d) = \Delta(d) = pr_\delta(d) = pr_{(\delta(d)\chi_d + \delta(b)\chi_b)}(d) = \delta(d)\cdot pr_d(d) + \delta(b)\cdot pr_b(d) \geq (\delta(d)+\delta(b))\cdot pr_d(d)$$

where in the last inequality we have used $pr_b(d) \leq pr_d(d)$ which is true by monotonicity. Then we continue

$$pr_a(d) \geq (\texttt{NetLeakage})\cdot pr_d(d) \geq \left(\frac{1-\alpha}{\alpha} c_1 p^* \phi(A) n\right)\cdot pr_d(d) \geq \left(\frac{1-\alpha}{\alpha} c_1 p^* \phi(A) n\right)\cdot\left(\frac{\pi\gamma}{2\ell} - O\Big(\frac{1}{\ell^2}\Big)\right)$$

using (A.6) in the second inequality and Lemma A.4 in the last inequality, so we conclude that $pr_a(d) \geq \frac{c_1}{2} p^* \phi(A) n(\ell - O(1))$ and then $\frac{pr_a(d)}{\deg(d)} \geq \frac{c_1 c_0}{2} p^*(1 - O(\frac{1}{\ell}))$. □

**Step 4: putting it all together.** We now define (using the fact that $\gamma > 0$ and $\gamma < 4$) constant $c_2$ to satisfy

$$1 - c_2 \stackrel{\text{def}}{=} \frac{1 - \frac{2\gamma}{\pi^2/4+\gamma} + \frac{2\gamma}{\pi^2+\gamma}}{1 - \frac{2\gamma}{\pi^2+\gamma}} < 1 \ .$$

This constant is asymptotically the ratio between (A.4) and (A.5), so once we let $p^* \stackrel{\text{def}}{=} \frac{pr_a(b)}{\deg(b)}$ it satisfies that (using the fact that $\phi(A)\ell^2 = o(1)$)

$$\frac{pr_a(c)}{\deg(c)} \leq (1-c_2)p^*(1+o(1)) \ .$$

Next, if we choose $c_1 = \frac{c_2}{2}$ and $c_0 = \frac{2}{c_1}$ in Lemma A.5, this gives

$$\frac{pr_a(d)}{\deg(d)} \geq \min\left\{1 - \frac{c_2}{2}, 1 - O\Big(\frac{1}{\ell}\Big)\right\} p^* \ .$$

It is now clear from the above two inequalities that in the asymptotic case, i.e., when $n, \ell$ are sufficiently large, we always have $\frac{pr_a(d)}{\deg(d)} > \frac{pr_a(c)}{\deg(c)}$. This finishes the proof of Lemma 5.1. ■

[14] This is obvious when $t = 0$. For $q^{(t+1)}$, we compute $p^{(t+1)} = p^{(t)}W$ and $q^{(t+1)} = q^{(t)}W + \delta^{(t)}$. Based on the inductive assumption that the claim holds for $q^{(t)}$, it is automatically true that for $u \in A \setminus \{b\}$, $p^{(t+1)}(u) = q^{(t+1)}(u)$, and $u \in V \setminus (A \cup \{d\})$ we have $q^{(t+1)}(u) = 0$. For $u = b$ or $u = d$, one can carefully check that $\delta^{(t)}$ is introduced to precisely make $q^{(t+1)}(b) = p^{(t+1)}(b)$ and $q^{(t+1)}(d) = 0$, so the claim holds.

# B Algorithm for Computing Approximate PageRank Vector

In this section we briefly summarize the algorithm `Approximate-PR` (see Algorithm 2) proposed by Andersen, Chung and Lang [ACL06] (based on the Jeh and Widom [JW03]) to compute an approximate PageRank vector. At high level, `Approximate-PR` is an iterative algorithm, and maintains an invariant that $p$ is always equal to $pr_{s-r}$ at each iteration.

Initially it lets $p = \vec{0}$ and $r = s$ so that $p = \vec{0} = pr_{\vec{0}}$ satisfies this invariant. Notice that $r$ does not necessarily satisfy $r(u) \le \varepsilon \deg(u)$ for all vertices $u$, and thus this $p$ is often not an $\varepsilon$-approximate PageRank vector according to Definition 2.2 at this initial step.

In each following iteration, `Approximate-PR` considers a vertex $u$ that violates the $\varepsilon$-approximation of $p$, i.e., $r(u) \ge \varepsilon \deg(u)$, and *pushes* this $r(u)$ amount of probability mass elsewhere:

- $\alpha \cdot r(u)$ amount of them is pushed to $p(u)$;
- $\frac{1-\alpha}{2\deg(u)} r(u)$ amount of them is pushed to $r(v)$ for each neighbor $v$ of $u$; and
- $\frac{1-\alpha}{2} r(u)$ amount of them remains at $r(u)$.

One can verify that after any push step the newly computed $p$ and $r$ will still satisfy $p = pr_{s-r}$. This indicates that the invariant is satisfied at all iterations. When `Approximate-PR` terminates, it satisfies both $p = pr_{s-r}$ and $r(u) \le \varepsilon \deg(u)$ for all vertices $u$, so $p$ must be an $\varepsilon$-approximate PageRank vector.

We are left to show that `Approximate-PR` terminates quickly, and the support volume of $p$ is small:

**Proposition 2.3.** *For any starting vector $s$ with $\|s\|_1 \le 1$ and $\varepsilon \in (0,1]$,* `Approximate-PR` *computes an $\varepsilon$-approximate PageRank vector $p = pr_{s-r}$ for some $r$ in time $O\left(\frac{1}{\varepsilon\alpha}\right)$, with* $\mathrm{vol}(\mathrm{supp}(p)) \le \frac{2}{(1-\alpha)\varepsilon}$.

*Proof sketch.* To show that this algorithm converges fast, one just needs to notice that at each iteration $\alpha r(u) \ge \alpha\varepsilon \deg(u)$ amount of probability mass is pushed from vector $r$ to vector $p$, so the total amount of them cannot exceed 1 (because $\|s\|_1 \le 1$). This gives $\sum_{i=1}^{T} \deg(u_i) \le \frac{1}{\varepsilon\alpha}$ where $u_i$ is the vertex chosen at the $i$-th iteration and $T$ is the number of iterations. However, it is not hard to verify that the total running time of `Approximate-PR` is exactly $O\big(\sum_{i=1}^{T} \deg(u_i)\big)$, and thus `Approximate-PR` runs in time $O\big(\frac{1}{\varepsilon\alpha}\big)$.

To bound the support volume, we consider an arbitrary vertex $u \in V$ with $p(u) > 0$. This $p(u)$ amount of probability mass must come from $r(u)$ during the algorithm, and thus vertex $u$ must be pushed at least once. Notice that when $u$ is lasted pushed, it satisfies $r(u) \ge \frac{1-\alpha}{2}\varepsilon \deg(u)$ after the push, and this value $r(u)$ cannot decrease in the remaining iterations of the algorithm. This implies that for all $u \in V$ with $p(u) > 0$, it must be true that $r(u) \ge \frac{1-\alpha}{2}\varepsilon \deg(u)$. However, we must have $\|r\|_1 \le 1$ because $\|s\|_1 \le 1$, so the total volume for such vertices cannot exceed $\frac{2}{(1-\alpha)\varepsilon}$. □

**Algorithm 2** `Approximate-PR` (from [ACL06])

**Input:** starting vector $s$, teleport probability $\alpha$, and approximate ratio $\varepsilon$.

**Output:** the $\varepsilon$-approximate PageRank vector $p = pr_{s-r}$.

```
p ← 0⃗ and r ← s.
while r(u) ≥ ε deg(u) for some vertex u ∈ V do
    Pick an arbitrary u satisfying r(u) ≥ ε deg(u).
    p(u) ← p(u) + αr(u).
    For each vertex v such that (u, v) ∈ E:
    r(v) ← r(v) + (1−α)/(2 deg(u)) r(u).
    r(u) ← (1−α)/2 r(u).
end while
return p.
```

## References

[ACE+13] L. Alvisi, A. Clement, A. Epasto, S. Lattanzi, and A. Panconesi. The evolution of sybil defense via social networks. In *IEEE Symposium on Security and Privacy*, 2013.

[ACL06] Reid Andersen, Fan Chung, and Kevin Lang. Using pagerank to locally partition a graph. 2006. An extended abstract appeared in FOCS '2006.

[AGM12] Reid Andersen, David F. Gleich, and Vahab Mirrokni. Overlapping clusters for distributed computation. WSDM '12, pages 273–282, 2012.

[AHK10] Sanjeev Arora, Elad Hazan, and Satyen Kale. O(sqrt(log(n)) approximation to sparsest cut in õ(n$^2$) time. *SIAM Journal on Computing*, 39(5):1748–1771, 2010.

[AK07] Sanjeev Arora and Satyen Kale. A combinatorial, primal-dual approach to semidefinite programs. STOC '07, pages 227–236, 2007.

[AL06] Reid Andersen and Kevin J. Lang. Communities from seed sets. WWW '06, pages 223–232, 2006.

[AL08] Reid Andersen and Kevin J. Lang. An algorithm for improving graph partitions. SODA, pages 651–660, 2008.

[Alo86] Noga Alon. Eigenvalues and expanders. *Combinatorica*, 6(2):83–96, 1986.

[AP09] Reid Andersen and Yuval Peres. Finding sparse cuts locally using evolving sets. STOC, 2009.

[ARV09] Sanjeev Arora, Satish Rao, and Umesh V. Vazirani. Expander flows, geometric embeddings and graph partitioning. *Journal of the ACM*, 56(2), 2009.

[AvL10] Morteza Alamgir and Ulrike von Luxburg. Multi-agent random walks for local clustering on graphs. ICDM '10, pages 18–27, 2010.

[CKK+06] Shuchi Chawla, Robert Krauthgamer, Ravi Kumar, Yuval Rabani, and D. Sivakumar. On the hardness of approximating multicut and sparsest-cut. *Computational Complexity*, 15(2):94–114, June 2006.

[GLMY11] Ullas Gargi, Wenjun Lu, Vahab S. Mirrokni, and Sangho Yoon. Large-scale community detection on youtube for topic discovery and exploration. In *AAAI Conference on Weblogs and Social Media*, 2011.

[GS12] David F. Gleich and C. Seshadhri. Vertex neighborhoods, low conductance cuts, and good seeds for local community methods. In *KDD '2012*, 2012.

[Hav02] Taher H. Haveliwala. Topic-sensitive pagerank. In *WWW '02*, pages 517–526, 2002.

[JW03] Glen Jeh and Jennifer Widom. Scaling personalized web search. In *WWW*, pages 271–279. ACM, 2003.

[KLL+13] Tsz Chiu Kwok, Lap Chi Lau, Yin Tat Lee, Shayan Oveis Gharan, and Luca Trevisan. Improved cheeger's inequality: Analysis of spectral partitioning algorithms through higher order spectral gap. In *STOC '13*, January 2013.

[KVV04] Ravi Kannan, Santosh Vempala, and Adrian Vetta. On clusterings: Good, bad and spectral. *Journal of the ACM*, 51(3):497–515, 2004.

[LC10] Frank Lin and William W. Cohen. Power iteration clustering. In *ICML '10*, pages 655–662, 2010.

[LLDM09] Jure Leskovec, Kevin J. Lang, Anirban Dasgupta, and Michael W. Mahoney. Community structure in large networks: Natural cluster sizes and the absence of large well-defined clusters. *Internet Mathematics*, 6(1):29–123, 2009.

[LLM10] Jure Leskovec, Kevin J. Lang, and Michael Mahoney. Empirical comparison of algorithms for network community detection. WWW, 2010.

[LR99] Frank Thomson Leighton and Satish Rao. Multicommodity max-flow min-cut theorems and their use in designing approximation algorithms. *Journal of the ACM*, 46(6):787–832, 1999.

[LR04] Kevin Lang and Satish Rao. A flow-based method for improving the expansion or conductance of graph cuts. *Integer Programming and Combinatorial Optimization*, 3064:325–337, 2004.

[LS90] László Lovász and Miklós Simonovits. The mixing rate of markov chains, an isoperimetric inequality, and computing the volume. FOCS, 1990.

[LS93] László Lovász and Miklós Simonovits. Random walks in a convex body and an improved volume algorithm. *Random Struct. Algorithms*, 4(4):359–412, 1993.

[Mer12] Mircea Merca. A note on cosine power sums. *Journal of Integer Sequences*, 15:12.5.3, May 2012.

[MMV12] Konstantin Makarychev, Yury Makarychev, and Aravindan Vijayaraghavan. Approximation algorithms for semi-random partitioning problems. In *STOC '12*, pages 367–384, 2012.

[MP03] Ben Morris and Yuval Peres. Evolving sets and mixing. STOC '03, pages 279–286. ACM, 2003.

[MR95] Rajeev Motwani and Prabhakar Raghavan. *Randomized algorithms.* Cambridge University Press, 1995.

[OSV12] Lorenzo Orecchia, Sushant Sachdeva, and Nisheeth K. Vishnoi. Approximating the exponential, the lanczos method and an $\tilde{O}(m)$-time spectral algorithm for balanced separator. In *STOC '12*. ACM Press, November 2012.

[OSVV08] Lorenzo Orecchia, Leonard J. Schulman, Umesh V. Vazirani, and Nisheeth K. Vishnoi. On partitioning graphs via single commodity flows. In *STOC 08*, New York, New York, USA, 2008.

[OT12] Shayan Oveis Gharan and Luca Trevisan. Approximating the expansion profile and almost optimal local graph clustering. FOCS, pages 187–196, 2012.

[OZ14] Lorenzo Orecchia and Zeyuan Allen Zhu. Flow-based algorithms for local graph clustering. SODA, 2014.

[Sch07] S. E. Schaeffer. Graph clustering. *Computer Science Review,*, 1(1):27–64, 2007.

[She09] Jonah Sherman. Breaking the multicommodity flow barrier for $o(\sqrt{\log n})$-approximations to sparsest cut. FOCS '09, pages 363–372, 2009.

[SJ89] Alistair Sinclair and Mark Jerrum. Approximate counting, uniform generation and rapidly mixing markov chains. *Information and Computation*, 82(1):93–133, 1989.

[SM00] J. Shi and J. Malik. Normalized cuts and image segmentation. *IEEE Transactions on Pattern Analysis and Machine Intelligence*, 22(8):888–905, 2000.

[SS08] Shai Shalev-Shwartz and Nathan Srebro. SVM optimization: inverse dependence on training set size. In *ICML*, 2008.

[ST04] Daniel Spielman and Shang-Hua Teng. Nearly-linear time algorithms for graph partitioning, graph sparsification, and solving linear systems. STOC, 2004.

[ST13] Daniel A. Spielman and Shang-Hua Teng. A local clustering algorithm for massive graphs and its application to nearly linear time graph partitioning. *SIAM Journal on Computing*, 42(1):1–26, January 2013.

[WLS+12] Xiao-Ming Wu, Zhenguo Li, Anthony Man-Cho So, John Wright, and Shih-Fu Chang. Learning with partially absorbing random walks. In *NIPS*, 2012.

[ZCZ+09] Zeyuan Allen Zhu, Weizhu Chen, Chenguang Zhu, Gang Wang, Haixun Wang, and Zheng Chen. Inverse time dependency in convex regularized learning. ICDM, 2009.

[ZLM13] Zeyuan Allen Zhu, Silvio Lattanzi, and Vahab Mirrokni. A local algorithm for finding well-connected clusters. In *ICML*, 2013. `http://jmlr.org/proceedings/papers/v28/allenzhu13.pdf`.